\documentclass[graybox]{svmult}

\usepackage{helvet}         % selects Helvetica as sans-serif font
\usepackage{courier}        % selects Courier as typewriter font
\usepackage{type1cm}        % activate if the above 3 fonts are
\usepackage{makeidx}         % allows index generation
\usepackage{graphicx}        % standard LaTeX graphics tool
\usepackage{multicol}        % used for the two-column index
\usepackage[bottom]{footmisc}% places footnotes at page bottom

\makeindex             % used for the subject index
\usepackage{enumitem}
\usepackage{amsmath,amssymb,mathrsfs}
\usepackage{mathtools}
\usepackage{stmaryrd}					% for \llbracket, \rrbracket
\DeclarePairedDelimiter\abs{\lvert}{\rvert}
\newcommand{\suchthat}{\mid}
\DeclareMathOperator{\card}{Card}
\newcommand{\R}{\mathbb{R}}
\newcommand{\partfun}{\mathcal{Z}}
\newcommand{\domain}{\mathcal{D}}
\newcommand{\macro}[1]{\mathcal{#1}} % Used for E[\omega],..
\newcommand{\meanfield}[1]{\mathscr{#1}}  % Used for S[\rho], E[\rho],...
\newcommand{\thermo}[1]{#1}		       % Used for thermodynamic potentials S(E,\Gamma_2,)...
\newcommand{\omegabf}{\boldsymbol{\omega}}

\begin{document}

\title*{An Introduction to Large Deviations and Equilibrium Statistical Mechanics for Turbulent Flows}
\titlerunning{Equilibrium Statistical Mechanics for Turbulent Flows}
\author{Corentin Herbert}
\institute{Corentin Herbert \at National Center for Atmospheric Research, Boulder, CO 80307, USA. \email{cherbert@ucar.edu}}
\maketitle

\abstract{Two-dimensional turbulent flows, and to some extent, geophysical flows, are systems with a large number of degrees of freedom, which, albeit fluctuating, exhibit some degree of organization: coherent structures emerge spontaneously at large scales. In this short course, we show how the principles of equilibrium statistical mechanics apply to this problem and predict the condensation of energy at large scales and allow for computing the resulting coherent structures. We focus on the structure of the theory using the language of large deviation theory.}

\section{Introduction}

Various characterizations of turbulent flows can be encountered; the components they usually entail are a chaotic dynamics on a strange attractor~\cite{RuelleBook1989}, a large range of scales (i.e. a large number of degrees of freedom), and strong nonlinear effects due to the prevalence of inertia over molecular dissipation~\cite{FalkovichBook}.
Such flows can be found in industrial problems, but also in nature, for instance in geophysical flows and astrophysical flows.
The above mentioned properties typically mean that not much can be said about the system in a deterministic framework, and that one should try instead to predict statistical properties. 

This is exactly the purpose of the field of statistical mechanics: given a dynamical system (or set of ordinary or partial differential equations) in a large phase space (the microscopic state), can we predict typical values for specific functions on phase space (the macroscopic observables) without knowing the exact trajectory in phase space? For a large class of systems, said to be \emph{in equilibrium}, such typical values can be obtained by assuming that the microscopic variables are random and distributed according to probability measures built upon a few macroscopic quantities, the invariants of the dynamical system.
A classical example is that of the ideal gas: the exact position and velocity of the molecules matters little to us, but knowing the relations between macroscopic quantities such as temperature, pressure, energy and entropy is fundamental.

The ideas of statistical mechanics have been applied successfully to a large number of models of physical phenomena. An example of achievement of this approach is the theory of phase transitions, in which systems such as the Ising model, a toy-model of ferromagnetism, have been instrumental. However, turbulent flows present a number of difficulties: (i) they are directly formulated as continuous fields (infinite number of degrees of freedom) and have an infinity of conserved quantities, (ii) the interactions between constituents have a long range, (iii) in many practical applications, the system is driven out of equilibrium by external forces. 

Although we shall not tackle issue (iii) at all in this chapter, we will try to show how (i) and (ii) are actually useful ingredients to make probabilistic predictions for the system. They are the cornerstones of a mean-field theory: interacting degrees of freedom can be treated as statistically independent random variables in the limit of a large number of degrees of freedom. A natural language to express these properties is that of \emph{large deviations theory}~\cite{LanfordBook,RuelleBook,EllisBook}: the probability of the outcome of a given observable concentrates exponentially around a set of values when the size of the system goes to infinity. The focus of the chapter is on the presentation of the large deviation principles for carefully chosen observables for a discretized form of 2D turbulence. To show that the principles at work are very general, we shall underline the connection with simpler models such as variants of the Ising model of ferromagnetism.
Although it is shown that the theory allows us to compute the equilibrium states of the system, we shall not dwell on the description of such equilibrium states; the reader is referred to the review articles~\cite{Bouchet2012,Lucarini2014} on this topic. We shall also refrain from discussing the connections with earlier applications of statistical mechanics, like the point vortex approach of Onsager, reviewed in~\cite{Eyink2006}, or the Kraichnan approach to Galerkin truncated flows~\cite{Kraichnan1980}, only mentioned briefly in section~\ref{kraichnansection}.

These notes are based on lectures given at the \emph{Stochastic Equations for Complex Systems: Theory and Applications} summer school organized at the University of Wyoming in June 2014. They mostly serve a pedagogical purpose, and we shall not give proofs of the results with the required mathematical rigor. However, we have tried as much as possible to provide the original references for the interested readers. The presentation adopted here owes much to the references~\cite{Touchette2009,Bouchet2010,Potters2013}.
Note that the ideas discussed here are applicable to many other systems with long range interactions~\cite{DauxoisLRIbook,Campa2009} and in particular gravitational systems~\cite{Padmanabhan1990,Chavanis2006h}, plasmas, cold atoms or toy models of statistical physics.

\section{Models of turbulent flows}

\subsection{3D and 2D hydrodynamics}

We are mainly interested here in the behavior of incompressible fluid flows, which is governed by the Navier-Stokes equations:
\begin{align}
\partial_t \vec{u} + \vec{u} \cdot \nabla \vec{u} &= - \nabla P + \nu \Delta \vec{u},\\
\nabla \cdot \vec{u} &=0,
\end{align}
where $\vec{u}$ is the velocity field, $P$ the pressure and $\nu$ the viscosity. The equations can be recast into non-dimensional form by introducing a velocity scale $U$, a length scale $L$, the corresponding time scale or \emph{eddy turnover time} $T=L/U$, and the Reynolds number $Re = UL/\nu$. In other words, the Reynolds number measures the ratio of the nonlinear term and the dissipative term, or equivalently, of inertia and viscosity~\cite{LandauFluidBook,FalkovichBook}. Since viscosity acts at small scales, it is also a measure of the range of scales characteristic of the flow: the smallest scale is the Kolmogorov scale $\ell_\eta=(\nu^3/\epsilon)^{1/4}$, where $\epsilon$ is the energy dissipation rate. Now, with $\epsilon=U^3/L$, we obtain $L/\ell_{\eta}=Re^{3/4}$. Hence, the effective number of degrees of freedom in 3D flows grows as $Re^{9/4}$: flows with Reynolds number on the order of $10^9$ are not uncommon in nature (the atmosphere and the ocean for instance), leading to a very large typical number of degrees of freedom.

The Navier-Stokes equations can be recast in terms of the vorticity field $\omegabf=\nabla \times \vec{u}$:
\begin{align}
\partial_t \omegabf + \vec{u} \cdot \nabla \omegabf &= \omegabf \cdot \nabla \vec{u} + \nu \Delta \omegabf.
\intertext{The first term on the right hand side corresponds to stretching of vorticity tubes. When we consider a flow on a two-dimensional surface rather than the three-dimensional space, this vorticity stretching term vanishes (the only non-vanishing component of vorticity $\omega$ is normal to the surface), yielding conservation of vorticity along streamlines in the inviscid ($\nu=0$) case:}
\partial_t \omega + \vec{u} \cdot \nabla \omega &= 0.\label{vort2deq}
\end{align}
This difference between 2D and 3D flows have important consequences on their respective behavior. While 3D flows tend to transfer energy from the large scales to the small scales, where it is dissipated by viscosity, in a process referred to as a direct energy cascade~\cite{Kolmogorov1941a,FrischBook} (big vortices break up into smaller and smaller vortices), 2D flows, on the contrary, transfer energy from the small scales to the large scales, and this is called an inverse energy cascade~\cite{Kraichnan1980,Sommeria2001b,Tabeling2002,Boffetta2012}. In this inverse cascade process, vortices merge to form larger and larger vortices~\cite{McWilliams1984}. Unless sufficient large scale dissipation (e.g. bottom friction) is present, the energy piles up at the largest available scales, forming a \emph{condensate} which dominates the flow~\cite{LSmith1993,Chertkov2007,Boffetta2012}.

The physical problem we are interested in here is the inverse energy cascade and the emergence of large scale coherent structures.

\subsection{Global invariants}

The equations of motion for 3D and 2D hydrodynamics have a Hamiltonian structure, although non-canonical: there exists a Poisson structure, but it is degenerate~\cite{OlverBook}. This degeneracy leads to the existence of invariants, described in this section.

Inviscid 3D flows have two global invariants, the energy and the helicity~\cite{Serre1984}:
\begin{align}
E &= \frac 1 2 \int \vec{u}^2,\\
H &= \int \vec{u} \cdot \omegabf.
\end{align}
Helicity being sign indefinite, it does not in general constrain the nonlinear transfers sufficiently to hamper the direct energy cascade process~\cite{Kraichnan1973} (see however,~\cite{Biferale2012,Herbert2014a} for particular cases). On the contrary, in 2D, vorticity conservation along streamlines leads to a family of invariants in addition to the energy ($\psi$ being the stream function, defined by $\omega=-\Delta \psi$)
\begin{align}
\macro{H} &= \int_\domain \omega(\vec{x})\psi(\vec{x})d\vec{x},
\intertext{the Casimir invariants:}
I_g &= \int_\domain g(\omega(\vec{x}))d\vec{x},
\end{align}
where $g$ is an arbitrary function. As a particular case, all the moments (or $L^p$ norms) of the vorticity field are conserved:
\begin{align}
\Gamma_p &= \frac 1 {\abs{\domain}} \int_\domain \omega(\vec{x})^p d\vec{x},
\end{align}
including the $L^2$ norm of the vorticity field, referred to as the \emph{enstrophy}. It was anticipated early on~\cite{Kraichnan1967,Leith1968,Batchelor1969} that the existence of a second, positive-definite, quadratic invariant, in addition to the energy, is sufficient to invert the direction of the energy cascade. The basic idea is that enstrophy is stronger in the presence of small-scale activity: transferring energy towards the small-scales while keeping the total energy constant cannot be done if we also need to conserve enstrophy. This loose statement was made more precise by a number of analytic arguments~\cite{Fjortoft1953,Kraichnan1967,Leith1968,Batchelor1969,Merilees1975,NazarenkoBook}, and verified in experiments~\cite{Paret1997,Rutgers1998} and high-resolution numerical simulations~\cite{Boffetta2012}. Statistical mechanics provides one of these analytical arguments (see section \ref{kraichnansection}).

Conservation of the Casimir invariants can be formulated equivalently in terms of the moments of the vorticity field, as above, or in terms of the vorticity distribution. Indeed, the fraction of the domain area $\abs{\domain}$, occupied by the vorticity level $\sigma$, which can be written as
\begin{align}
\gamma(\sigma) &= \frac 1 {\abs{\domain}} \int_\domain \delta(\omega(\vec{x})-\sigma)d\vec{x},
\end{align}
is conserved. We shall see that this form is particularly convenient in section \ref{meanfieldsection}, but note that the two formulations are connected by the formula
\begin{align}
\Gamma_p &= \frac 1 {\abs{\domain}}\int_\domain d\vec{x}\int_\R d\sigma \sigma^p \delta(\omega(\vec{x})-\sigma) = \int_\R d\sigma \sigma^p \gamma(\sigma),\label{momvorteq}
\intertext{and, conversely, using an integral representation of the Dirac distribution,}
\gamma(\sigma) &= \frac{1}{2\pi} \sum_{p=0}^{+\infty} \frac{(-1)^p \Gamma_p}{p!} \delta^{(p)}(\sigma).
\end{align}
Finally, note that the vorticity distribution is normalized:
\begin{align}
\int_{\R} \gamma(\sigma)d\sigma &=1.\label{gammanormeq}
\end{align}

\subsection{Geophysical flows}

Although 2D flows are interesting in themselves, part of the motivation for studying them comes from their common features with geophysical flows. Indeed, in addition to the small aspect ratio of the atmosphere and the ocean, their dynamics is subjected to the effect of strong rotation and density stratification. These properties allow for an asymptotic regime which describes well the large-scale dynamics, the \emph{quasi-geostrophic} regime~\cite{VallisBook}. This regime is very similar to 2D flows, because it reduces to a quantity, called \emph{potential vorticity}, being advected by the flow, similarly to the vorticity (see~\eqref{vort2deq}). In particular, the velocity field is purely horizontal. The only difference is that the fields also depend on the vertical, and whereas the vorticity is the laplacian of the stream function: $\omega=-\Delta \psi$ in 2D, here the potential vorticity is related to the stream function by a slightly more complicated linear differential operator.
The existence of Casimir invariants similar to those of 2D flows leads again to an inverse cascade of energy and the formation of coherent structures at large scales~\cite{Charney1971,Rhines1979,SalmonBook}. Therefore, the considerations presented here may apply to such flows as well, and attempts to extend the theory in this context have flourished over the past few years. For the sake of simplicity, we shall restrict ourselves here to the case of 2D flows; the interested reader may consult the literature on extensions to quasi-geostrophic flows in the barotropic case~\cite{Bouchet2002,Naso2011,Venaille2011b,Herbert2012b}, the baroclinic case~\cite{DiBattista2001a,Venaille2012a,Venaille2012b,Herbert2014b}, shallow-water equations~\cite{Chavanis2002a,Chavanis2006b}, as well as the general references~\cite{MajdaWangBook,Bouchet2012,Lucarini2014}, for instance.

The quasi-geostrophic regime breaks down at smaller scales, and we enter an intermediate regime, often referred to as stratified turbulence~\cite{Lilly1983}. In this regime, even though we can still define a potential vorticity which is a Lagrangian invariant, it does not put as strong a constraint on the system as in the 2D case. Indeed, the fields (velocity, density) can be decomposed into a \emph{balanced} part which contributes to potential vorticity, and \emph{inertia-gravity waves}, which do not. As a result, the organization of the system in terms of inertial ranges and energy cascades is not so simple. High-resolution numerical simulations have indicated the existence of two inertial ranges with a constant and opposite flux of energy~\cite{Pouquet2013}. A possible interpretation is that the vortical modes are responsible for the inverse cascade of energy while the inertia-gravity waves have to do with the direct energy cascade. This interpretation is supported by a statistical mechanics argument~\cite{Herbert2014c}, which is an adaptation of the Kraichnan argument (see section~\ref{kraichnansection}) in the context of the restricted canonical ensemble~\cite{Penrose1979}.

Independently of the constraining effect of rotation and stratification (which can be seen as forces breaking isotropy), another direction of generalization which has been considered is that of 3D flows with symmetries, and especially axisymmetric flows~\cite{Leprovost2006,Naso2010b,Thalabard2014}. This configuration is relevant for setups used in laboratory experiments, such as the von Karman experiment. It has been shown in particular that one could define a microcanonical measure using an approach analogous to that of section \ref{microcanomeasuresection}, with, however, some considerable complications to treat the fluctuations of the poloidal field~\cite{Thalabard2014}.

\subsection{Discretized form for 2D Euler flows and analogies with toy models of magnetic systems}\label{discretizationsec}

\begin{figure}
\includegraphics[width=\linewidth]{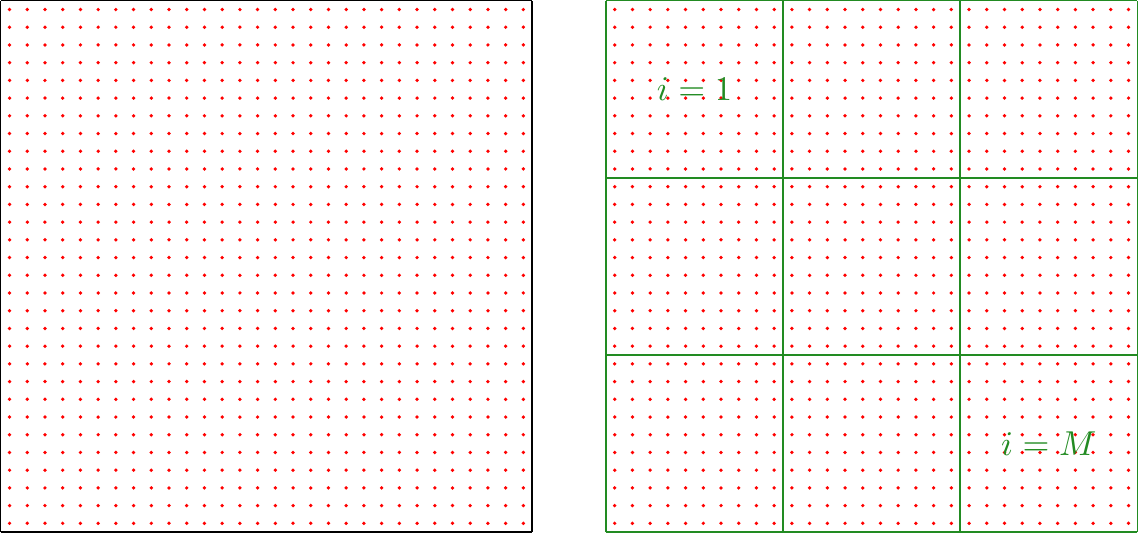}
\caption{Discretization and coarse-graining. Left: We replace the 2D domain by a finite lattice and the continuous vorticity field by a $N$-dimensional vector whose components are the values of the vorticity at each site. Right: We decompose the lattice in $M$ cells, each containing $n=N/M$ sites. The coarse-grained vorticity is a $M$ dimensional vector whose components are the average value of the vorticity in each cell.}\label{latticefig}
\end{figure}

Instead of the continuous vorticity field $\omega$ and the infinite dimensional phase space it belongs to, it may be more convenient to introduce finite dimensional models. Here we shall mostly consider a discretization on a square lattice with $N$ sites equally spaced in the domain $\domain$ (see Fig.~\ref{latticefig}), and the variables of interest are the values taken by vorticity at each site. 
In this form, the system can be related to some classical models of statistical physics.

\subsubsection{Two-vorticity level system and long-range Ising model}\label{isingmfsec}
The Ising model is one of the most famous models in statistical physics. It can be seen as a toy model of ferromagnetism, but it has served as a testbed for a very large number of ideas going far beyond this particular problem~\cite{Cipra1987}. 
It consists of a finite number $N$ of spins $s_i \in \{ -1,1\}$ located on a lattice of arbitrary shape and dimension (although a square lattice is often considered) and interacting through a hamiltonian of the form:
\begin{align}
\mathcal{H}_I[\hat{s}] &=- \frac 1 N \sum_{i,j=1}^N J_{ij}s_is_j.
\end{align}
In this form, the hamiltonian is just any quadratic function. A standard choice of interaction is the nearest-neighbor model: $J_{ij}=J$ if the sites $i$ and $j$ are connected in the lattice, and $J_{ij}=0$ if they are not. That way, aligned neighboring spins will contribute a term $-J$ to the hamiltonian, while anti-aligned neighboring spins will contribute $J$. If $J$ is positive the system is called \emph{ferromagnetic} and if it is negative the system is called \emph{antiferromagnetic}. An observable of interest is the \emph{magnetization}:
\begin{align}
\mathcal{M}[\hat{s}] &= \frac 1 N \sum_{i=1}^N s_i.
\end{align}
When one finds about the same proportion of positive and negative spins, the magnetization should vanish. Applying an external magnetic field, represented by a term of the form $-h\sum_i s_i$ in the hamiltonian, leads to alignment of spins, and therefore a non-vanishing magnetization. This is the standard behavior of so-called \emph{paramagnetic} materials. By contrast, in \emph{ferromagnetic} materials, spins may align spontaneously and yield unit magnetization (in absolute value) without imposing an external magnetic field (or, in experiments, the system retains its magnetization when the applied magnetic field is switched off). The Ising model can be seen as a toy model of the paramagnetic-ferromagnetic transition.

In the above case, the system has short-range interactions, since only neighboring spins interact. Versions with long-range interactions can be built by allowing non-vanishing $J_{ij}$ for distant sites $i$ and $j$. For instance, one may assume that all the spins interact with all the other spins with the same coupling constant: $J_{ij}=1/N$, where the $1/N$ ensures that the Hamiltonian is an intensive quantity. In this case, the Hamiltonian becomes a function of magnetization only:
\begin{align}
\mathcal{H}_{IMF}[\hat{s}] &= -\frac 1 {N^2} \sum_{i,j=1}^N s_i s_j = -\mathcal{M}[\hat{s}]^2.
\end{align}
This version of the Ising model is referred to as \emph{mean-field}, because it is tantamount to saying that each spins feels the effect of a magnetic field created by all the other spins rather than the individual effect of each of his neighbors. Indeed, let us consider a given spin $s_i$; it provides a contribution $-1/N s_i \sum_{j} J_{ij} s_j$, which is the same has a non-interacting spin under external magnetic field $1/N \sum_j J_{ij} s_j$ would. If we replace this magnetic field by the magnetization, we obtain the mean-field Hamiltonian. Note that the geometric shape (square, triangle, etc) and the dimension of the lattice do not matter here since all the spins interact with the same intensity.

An advantage of the mean-field Ising model is that it has an exact solution in any dimension~\cite{BaxterBook}. On the contrary, exact solutions for the standard, short-range Ising models are only know for dimension one~\cite{Ising1925} and two~\cite{Onsager1944}.

The discretized version of 2D flows described above is related to the Ising model in the following way: rather than allowing the vorticity to take any real value, we can restrict it to a two-level set $\{\sigma,-\sigma\}$.
Then the system becomes analogous to the Ising model, with an interaction matrix given by the Green function of the Laplacian on the lattice. On a plane, this amounts to interactions proportional to the logarithm of the distance between sites: $J_{ij} \propto \ln \abs{i-j}$ for $i \neq j$. This is a kind of long-range interaction. The difference with the Ising model is the presence of the vorticity distribution conservation constraint. This would amount to fixing the number of $+$ spins and the number of $-$ spins in the Ising model.

\subsubsection{Energy-enstrophy ensemble and the long-range spherical model}\label{sphericalmodelsec}

Another variant of the Ising model consists in letting the spins $s_i$ take any real value, while satisfying the global constraint $\sum_{i=1}^N s_i^2 =N$. Clearly, this constraint is satisfied in the standard Ising model with spins in $\{1,-1\}$. The name \emph{spherical model} was coined for this variant because of the form of the global constraint, which means that the set of all spin values lies on the surface of a sphere in $\R^{N}$. It was introduced by Berlin and Kac~\cite{Berlin1952} as an attempt to patch the divergence arising from assuming that the spins are distributed according to a normal distribution (the \emph{Gaussian model}) while remaining exactly solvable in any dimension~\cite{BaxterBook}. 
The observables of interest (Hamiltonian, magnetization) are the same as for the Ising model. Versions with short-range~\cite{Berlin1952} or long-range~\cite{Joyce1966} interactions can again be considered by choosing different quadratic forms $J_{ij}$.

In their discretized version, 2D flows resemble a long-range spherical model if we only retain one Casimir invariant: the enstrophy. Indeed, enstrophy conservation implies $\sum_i \omega_i^2 = N\Gamma_2$. Again, the interaction matrix is given by the Green function of the Laplacian on the lattice. This connection is further investigated in section~\ref{energyenstrophysection}. It has also been pointed out in a series of papers by Lim~\cite{Lim2001c,Lim2012}.

\section{Mean-field theory for 2D flows}\label{meanfieldsection}

We provide here a heuristic presentation of the mean-field theory introduced independently by Miller~\cite{Miller1990,Miller1992}, Robert and Sommeria~\cite{Robert1991a,Robert1991b}, and further developed by many others. The presentation is inspired by the original work by Miller and the more recent references~\cite{Bouchet2010,Bouchet2012,Potters2013}. More rigorous mathematical proofs can be found in the original papers by Robert and coworkers~\cite{Robert1989,Robert1990,Robert1991a,Robert1991b,Michel1994a,Michel1994b,Robert2000} and Ellis, Turkington and coworkers~\cite{Turkington1999,Boucher1999,Boucher2000,Ellis2000,Ellis2002}. 

\subsection{Microcanonical measure and large deviations for the energy and vorticity distribution}\label{microcanomeasuresection}

The general idea is to consider the vorticity field $\omega$, referred to as the \emph{microstate}, as a random variable distributed according to the \emph{microcanonical distribution}. In other words, we introduce a probability measure on the phase space $\Lambda=L^\infty(\domain)$, where $\domain$ is a 2D domain (we shall mostly consider the case of a rectangular domain here). We are going to give a sketch of the construction of this measure as a limit of measures on finite-dimensional phase spaces corresponding to approximations of the continuous vorticity field. Then, we will be able to make predictions on the value of \emph{macrostates}, i.e. observables $\macro{A}: \Lambda \longrightarrow \R$ (or more generally $\macro{A}: \Lambda \longrightarrow M$ where the space $M$ is \emph{macroscopic} in some sense, e.g. has a dimension much lower than $\Lambda$) on phase space, which, as we shall see, satisfy large deviation properties: they concentrate in probability around some specific values, the equilibrium states.

To keep things simple, we shall consider a finite number of vorticity levels $\mathfrak{S}=\{\sigma_1,\ldots,\sigma_K\}$. This amounts to saying that the vorticity distribution has the form $\gamma(\sigma)= \sum_{k=1}^{K} \gamma_k \delta(\sigma-\sigma_k)$.
We consider the discretized system with $N$ sites on the square lattice introduced in section \ref{discretizationsec} (see Fig.~\ref{latticefig}), and define a \emph{microstate} as being simply the value of the vorticity field at all the points of the lattice. Therefore the phase space is simply $\Lambda_N=\mathfrak{S}^N$.

Considering the conservation laws mentioned above, there are two observables of primary interest: the energy observable, i.e. the Hamiltonian, given by
\begin{align}
&\macro{H}_N: \hat{\omega} \in \Lambda_N \longmapsto \frac 1 {2N^2} \sum_{1 \leq i \neq j \leq N} G_{ij}^N \omega_{i} \omega_{j},\\
\intertext{with $G_{ij}^N$ the Green function of the Laplacian on the lattice, and the vorticity distribution observables}
&\macro{G}^{(k)}_N: \hat{\omega} \in \Lambda_N \longmapsto  \frac 1 N  \sum_{i=1}^N \delta_{\omega_{i},\sigma_k}.
\end{align}
Note that the set of accessible energies (i.e. the values taken by the observable $\macro{H}_N$) is finite and depends both on the vorticity levels $\sigma_k$ and on the number of sites $N$. Ultimately, in the limit $N \to \infty$, we shall be interested in a continuum of energy levels. One approach to circumvent this difficulty is to consider in a first step energy shells with finite width $\Delta E$, large enough so that each shell is attained by the energy observable for some microstates~\cite{Potters2013}. In the limit $N \to +\infty$, the results will not depend on the value of $\Delta E$. To keep notations as simple as possible, we shall refrain from doing so here, but in all rigor one should understand $\macro{H}_N[\hat{\omega}] \in [E,E+\Delta E]$ whenever we write $\macro{H}_N[\hat{\omega}]=E$.
In this framework, the set of microstates with vorticity distribution $\gamma$ and energy $E$ is
\begin{align}
&\Lambda_N(\gamma,E) = \{ \hat{\omega} \in \Lambda_N \suchthat \macro{H}_N[\hat{\omega}] = E, \forall k \in \llbracket 1, K \rrbracket, \macro{G}^{(k)}_N[\hat{\omega}]=\gamma_k \},\\
&= \macro{H}_N^{-1}(\{E\}) \cap \bigcap_{k=1}^K {\macro{G}_N^{(k)}}^{-1}(\{\gamma_k\}).
\end{align}
This is a finite set whose cardinality we denote by $\Omega_N(\gamma,E) = \card \Lambda_N(\gamma,E)$.

We are going to introduce two probability measures on phase space: first, let us consider a \emph{prior measure} $\mu^{(N)}$, which here is just the normalized counting measure: if $M \subset \Lambda_N, \mu^{(N)}(M)=\frac{\card M}{K^N}$. This amounts to saying that all the microstates are equiprobable: for any observable $\macro{A}_N: \Lambda_N \longrightarrow \R$, the probability of the outcome $x$ is just the fraction of microstates for which $\macro{A}_N[\hat{\omega}]=x$. Now, we want to restrict that statement to all the microstates with a fixed energy and vorticity distribution, while assigning vanishing probability to all the other microstates. Hence, we introduce the (finite-$N$) microcanonical measure $\mu_{\gamma,E}^{(N)}$: if $M \subset \Lambda_N$, $\mu_{\gamma,E}^{(N)}(M)=\card(M \cap \Lambda_N(\gamma,E))/\Omega_N(\gamma,E)$. Hence, for an observable $\macro{A}_N$, the probability law of the random variable $\macro{A}_N[\hat{\omega}]$ is $\mu_{\gamma,E}^{(N)}(\macro{A}_N[\hat{\omega}]=x)=\mu_{\gamma,E}^{(N)}(\macro{A}_N^{-1}(\{x\}))$. Note that we have introduced indices $\gamma$ and $E$ to distinguish from probabilities computed with respect to the prior measure. Probabilities in the microcanonical ensemble are thus just conditional probabilities:
\begin{align}
\mu_{\gamma,E}^{(N)}(\macro{A}_N[\hat{\omega}]=x) = \mu^{(N)}(\macro{A}_N[\hat{\omega}]=x | \macro{H}_N[\hat{\omega}]=E, \macro{G}_N^{(k)}[\hat{\omega}]=\gamma_k),\\
= \begin{cases}
\frac{\mu^{(N)}(\macro{A}_N[\hat{\omega}]=x,\macro{H}_N[\hat{\omega}]=E, \macro{G}_N^{(k)}[\hat{\omega}]=\gamma_k)}{\mu^{(N)}(\macro{H}_N[\hat{\omega}]=E, \macro{G}_N^{(k)}[\hat{\omega}]=\gamma_k))} & \substack{\text{ if } \macro{H}_N[\hat{\omega}]=E, \text{ and }\\ \forall k \in \llbracket 1, K\rrbracket, \macro{G}_N^{(k)}[\hat{\omega}]=\gamma_k}\\
0 & \text{ otherwise.}
\end{cases}
\end{align}

As mentioned above, observables of particular interest are the hamiltonian $\macro{H}_N$ and the vorticity distribution observables $\macro{G}_N^{(k)}$.
The joint probability to observe an energy $E$ and a vorticity distribution $\gamma$, with respect to the prior measure, satisfies a large-deviation property, and the large deviation rate function is the opposite of the entropy $S(E,\gamma)$:
\begin{align}
\mu^{(N)}(\macro{H}_N[\hat{\omega}] = E, \macro{G}_N^{(k)}[\hat{\omega}]=\gamma_k) = \frac{\Omega_N(\gamma,E)}{K^N} = e^{N S(E,\gamma) + o(N)},
\end{align}
with 
\begin{align}
S(E,\gamma)=\lim_{N \to \infty} \frac 1 N \ln \Omega_N(\gamma,E). \label{boltzmanentropyeq}
\end{align}

\subsection{Large deviations for the macrostates}\label{meanfieldsec}

\begin{figure}
\centering
\includegraphics[width=\linewidth]{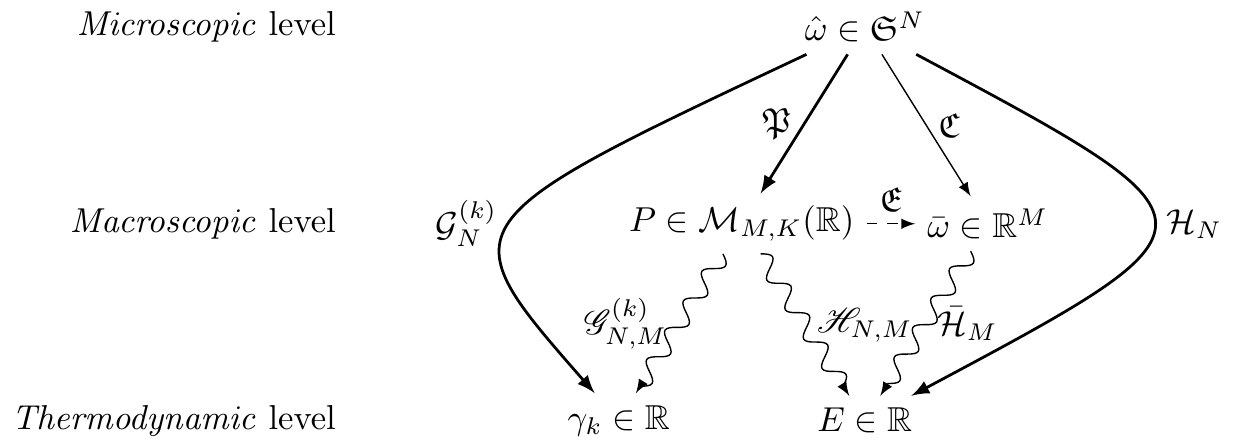}
\caption{The different levels of description of the system, and the observables/representation function relating them. Observables are represented with straight arrows, representation functions with wiggly arrows and contraction principles with dashed arrows. Observables for which a large deviation principle has been obtained directly are represented with thick arrows.}\label{micromacrothermofig}
\end{figure}

We now introduce a new class of observables associated with the coarse-graining of the vorticity field. We decompose the lattice into $M$ cells, each containing $n=N/M$ sites. For a microstate $\hat{\omega} \in \mathfrak{S}^N$, we shall denote the components as $\omega_{i\alpha}$ where $1 \leq i \leq M$ is the index of the cell and $1 \leq \alpha \leq n$ is the index of the site within the cell (see Fig.~\ref{latticefig}). The coarse-graining observable is given by
\begin{align}
\mathfrak{C}: \hat{\omega} \in \mathfrak{S}^N \longmapsto \bar{\omega} \in \R^M, \text{ with } \bar{\omega}_i=\frac 1 n \sum_{\alpha=1}^n\omega_{i\alpha}, 1\leq i \leq M.
\end{align}
More generally, we can define an observable which corresponds to the distribution of vorticity levels in each cell. It is just the empirical vector
\begin{align}
\mathfrak{P}: \hat{\omega} \in \mathfrak{S}^N \longmapsto P=(p_{ik})_{\substack{1 \leq i \leq M \\ 1 \leq k \leq K}} \in \mathcal{M}_{M,K}(\R), \text{ with } p_{ik}=\frac 1 n \sum_{\alpha=1}^n \delta_{\omega_{i\alpha},\sigma_k}.
\end{align}
Note that $\sum_{k=1}^K p_{ik}=1$. Besides, the observable $\mathfrak{C}$ can be deduced from $\mathfrak{P}$ since $\bar{\omega}_i=\sum_{k=1}^K \sigma_k p_{ik}$ for $1 \leq i \leq M$. Let us refer to the elements of the image of $\mathfrak{P}$ as the \emph{macrostates}. The set of microstates corresponding to a given macrostate $P$ is simply its pre-image $\mathfrak{P}^{-1}(P)$. The number of microstates realizing a given macrostate $P$ will be denoted $W(P)=\card \mathfrak{P}^{-1}(P)$. It is easily computed that:
\begin{align}
W(P)&=\prod_{i=1}^M \binom{n}{np_{i1}}\binom{n-np_{i1}}{np_{i2}}\cdots \binom{n-np_{i1}-\cdots-np_{in-1}}{np_{in}},\\
&=\prod_{i=1}^M \frac{n!}{\prod_{k=1}^K (n p_{ik})!}.
\end{align}
The vorticity distribution observables $\macro{G}_N^{(k)}$ take a constant value over an equivalence class $\mathfrak{P}^{-1}(P)$:
\begin{align}
\macro{G}_N^{(k)}[\hat{\omega}] &= \frac 1 M \sum_{i=1}^M p_{ik},
\end{align}
so that if $\hat{\omega}_1,\hat{\omega}_2 \in \mathfrak{S}^N$ are such that $\mathfrak{P}[\hat{\omega}_1]=\mathfrak{P}[\hat{\omega}_2]$, then for $1 \leq k \leq K$, $\macro{G}_N^{(k)}[\hat{\omega}_1]=\macro{G}_N^{(k)}[\hat{\omega}_2]$. In other words, the equivalence kernel of the observable $\mathfrak{P}$ is finer than that of any of the observables $\macro{G}_N^{(k)}$. In practice, this means that we need not worry about enforcing the vorticity distribution constraint when counting the number of microstates realizing a given macrostate.
For the energy observable, the situation is slightly more subtle: denoting $G_{i\alpha,j\beta}^{M,n}$ the Green function of the Laplacian on the lattice with the new indexing of the sites, the energy observable is given by:
\begin{align}
\macro{H}_{N,M}[\hat{\omega}] &= \frac 1 {2N^2} \sum_{\substack{1 \leq i,j \leq M\\ 1 \leq \alpha,\beta \leq n \\ (i,\alpha) \neq (j,\beta)}} G_{i\alpha,j\beta}^{M,n} \omega_{i\alpha}\omega_{j\beta},
\intertext{which is not necessarily constant over equivalence classes. However, it can be shown that the dominant contribution is the \emph{mean-field} energy, i.e. the energy of the coarse-grained vorticity field:}
\macro{H}_{N,M}[\hat{\omega}] &= \frac 1 {2M^2} \sum_{1 \leq i \neq j \leq M} G_{ij}^M \bar{\omega}_{i}\bar{\omega}_{j}+ o\left(\frac 1 N \right),\\
&= \macro{H}_M[\mathfrak{C}[\hat{\omega}]] + o\left(\frac 1 N \right).
\end{align}
The above results are sometimes restated by saying that we have an \emph{energy} (and here, also vorticity distribution) \emph{representation function}~\cite{Touchette2009} (see Fig.~\ref{micromacrothermofig}). It allows us to obtain the most probable states with respect to the microcanonical measure by obtaining a large deviation property with respect to the prior (unconstrained) measure.

Indeed, the unconstrained probability of observing a macrostate $P$ is
\begin{align}
\mu^{(N)}(\mathfrak{P}[\hat{\omega}]=P) &= \mu^{(N)}(\mathfrak{P}^{-1}(P)),\\ 
&= \frac{W(P)}{K^N}.
\intertext{Using the Stirling approximation, it is easily shown that when $N\to \infty$, this probability satisfies a large deviation property:}
\mu^{(N)}(\mathfrak{P}[\hat{\omega}]=P) &= e^{N\meanfield{S}_{M,K}[P]+o(N)},\label{macrostatelargedeveq}
\intertext{where we have introduced the mean-field entropy}
\meanfield{S}_{M,K}[P] &= \lim_{N \to +\infty} \frac 1 N \ln \mu^{(N)}(\mathfrak{P}[\hat{\omega}]=P),\\
&=- \frac 1 M \sum_{i=1}^M \sum_{k=1}^K p_{ik} \ln p_{ik}, \label{meanfieldentropyeq}
\end{align}
which again appears as a large deviation rate function (up to an additive constant and a minus sign), although this time it is a large deviation of an empirical vector (observable $\mathfrak{P}$) rather than a sample mean (energy observable $\macro{H}$). Hence, the above result should in all rigor be seen as a consequence of the Sanov theorem.

Now, in the microcanonical ensemble, the probability $\mu_{\gamma,E}^{(N)}(\mathfrak{P}[\hat{\omega}]=P)$ involves the joint (unconstrained) probability $\mu^{(N)}(\mathfrak{P}[\hat{\omega}]=P,\macro{H}_N[\hat{\omega}]=E,\macro{G}_N^{(k)}[\hat{\omega}]=\gamma_k)$. But due to the existence of the energy and vorticity distribution representation functions, we have:
\begin{equation}
\begin{split}
\mu^{(N)}(\mathfrak{P}[\hat{\omega}]=P,&\macro{H}_N[\hat{\omega}]=E,\macro{G}_N^{(k)}[\hat{\omega}]=\gamma_k) \\
&= \mu^{(N)}(\mathfrak{P}[\hat{\omega}]=P, \meanfield{H}_{N,M}[P]=E,\meanfield{G}_{N,M}^{(k)}[P]=\gamma_k),
\end{split}
\end{equation}
therefore,
\begin{align}
\mu_{\gamma,E}^{(N)}(\mathfrak{P}[\hat{\omega}]=P)&= \begin{cases}
\frac{\mu^{(N)}(\mathfrak{P}[\hat{\omega}]=P)}{\mu^{(N)}(\macro{H}_N[\hat{\omega}]=E, \macro{G}_N^{(k)}[\hat{\omega}]=\gamma_k))} & \substack{\text{ if } \meanfield{H}_{N,M}[P]=E, \text{ and }\\ \forall k \in \llbracket 1, K\rrbracket, \meanfield{G}_{N,M}^{(k)}[P]=\gamma_k}\\
0 & \text{ otherwise.}
\end{cases}
\end{align}
It follows that the probability of a given macrostate also satisfies a large deviation result with respect to the microcanonical measure:
\begin{align}
\mu_{\gamma,E}^{(N)}(\mathfrak{P}[\hat{\omega}]=P) &= e^{-NI[P]+o(N)},
\end{align}
with the large deviation rate function
\begin{align}
I[P]&=\begin{cases}
S(E,\gamma)- \meanfield{S}_{M,K}[P] & \text{ if } \meanfield{H}_{N,M}[P]=E, \text{ and } \forall k \in \llbracket 1, K\rrbracket, \meanfield{G}_{N,M}^{(k)}[P]=\gamma_k\\
+\infty & \text{ otherwise.} 
\end{cases}
\end{align}
Hence, the most probable macrostates with respect to the microcanonical measure are those which minimize the large deviation rate function, i.e. those which maximize the entropy $\meanfield{S}_{M,K}$ while satisfying the constraints on energy and vorticity distribution: they are solutions of a constrained variational problem.
It is worthy of note that the Boltzmann-Gibbs entropy $\meanfield{S}_{M,K}$, defined in~\eqref{meanfieldentropyeq}, evaluated at a solution $P^*$ of the variational problem, agrees with the entropy $S(E,\gamma)$ defined from the Boltzman formula~\eqref{boltzmanentropyeq}. This is not a coincidence, but a cornerstone of the mean-field approach. It can be understood in the language of large deviation theory as a \emph{contraction principle}~\cite{Touchette2009}. Roughly speaking, due to the existence of representation functions, the probability of observing an energy $E$ and a vorticity distribution $\gamma$ can be computed as the integral over all the macrostates (rather than the microstates) with these constraints: denoting
\begin{align}
\tilde{\Lambda}_{N,M}(\gamma,E) = \{ P \in \mathcal{M}_{M,K}(\R) \suchthat \meanfield{H}_{N,M}[P] = E, \forall k \in \llbracket 1, K \rrbracket, \meanfield{G}^{(k)}_{N,M}[P]=\gamma_k \},
\end{align}
we have
\begin{align}
\mu^{(N)} (\macro{H}_N[\hat{\omega}] = E, \macro{G}_N^{(k)}[\hat{\omega}]=\gamma_k) &= \int_{\Lambda_{N}(\gamma,E)} \mu^{(N)}(d\hat{\omega}),\\
&=\int_{\tilde{\Lambda}_{N,M}(\gamma,E)} \mu^{(N)}(\mathfrak{P}^{-1}(P)),\\
&= \int_{\tilde{\Lambda}_{N,M}(\gamma,E)} e^{N\meanfield{S}_{M,K}[P]+o(N)},\\
\intertext{Using Laplace's approximation, the integral evaluates to}
&= \exp\left( N\max_{P \in \tilde{\Lambda}_{N,M}(\gamma,E)}\meanfield{S}_{M,K}[P] + o(N)\right).
\end{align}
As a conclusion, the most probables macrostates $P^*$ with respect to the microcanonical measure satisfy $I[P^*]=0$: they are solutions of the constrained variational problem:
\begin{align}
S(\thermo{E},\gamma) = \max_{P} \{ \meanfield{S}_{M,K}[P] \suchthat \meanfield{H}_{N,M}[P]=\thermo{E}, \forall k \in \llbracket 1,K\rrbracket, \meanfield{G}_{N,M}^{(k)}[P]=\gamma_k\}.\label{mcvpeq}
\end{align}

\subsection{Thermodynamic limit and mean-field equation}\label{meanfieldeqsection}

We are now interested in the macrostates obtained in the limit $M \to +\infty$. Letting also $K \to +\infty$, they are the probability distributions for fine-grained vorticity $\rho(\vec{r},\sigma)$: $\rho(\vec{r},\sigma)d\sigma$ is the probability that the vorticity at point $\vec{r}$ lies in the interval $[\sigma,\sigma+d\sigma]$. The local normalization condition $\int_\R \rho(\vec{r},\sigma)d\sigma=1$ must still be satisfied for each point $\vec{r} \in \domain$. The coarse-grained vorticity field is now $\bar{\omega}(\vec{r})=\int_\R \sigma \rho(\vec{r},\sigma)d\sigma$. 
As explained above, the energy and vorticity distribution depend only on the macrostate $\rho$:
\begin{align}
\meanfield{H}[\rho] &= \frac 1 2 \int_{\domain^2} d\vec{r}d\vec{r'} \int_{\R^2} d\sigma d\sigma' G(\vec{r},\vec{r'}) \sigma \sigma'\rho(\vec{r},\sigma) \rho(\vec{r'},\sigma'),\\
\meanfield{D}_\sigma[\rho] &= \int_\domain \rho(\sigma,\vec{r})d\vec{r}.
\intertext{Similarly, the mean field entropy becomes}
\meanfield{S}[\rho] &= - \int_\domain d\vec{r} \int_\R d\sigma \rho(\sigma,\vec{r}) \ln \rho(\sigma,\vec{r}).\label{mfentropyeq}
\end{align}
The most probable macrostates are now those maximizing \eqref{mfentropyeq} while satisfying the energy and vorticity distribution constraints. They are solutions of the microcanonical variational problem:
\begin{align}
S(\thermo{E},\gamma) &= \max_{\rho} \{ \meanfield{S}[\rho] \suchthat \meanfield{H}[\rho]=\thermo{E}, \forall \sigma \in \R, \meanfield{D}_\sigma[\rho]=\gamma(\sigma)\}.
\end{align}
The critical points of the variational problem are readily found: there exist Lagrange multipliers $\beta$ and $\alpha(\sigma)$ such that the first variations vanish:
\begin{align}
0 &=\delta \meanfield{S} - \int_\domain d\vec{r} \zeta(\vec{r}) \int_\R d\sigma\delta \rho(\sigma,\vec{r}) - \beta \delta \meanfield{H} - \int_\R d\sigma \alpha(\sigma) \int_\domain d\vec{r}\delta \rho(\sigma,\vec{r}),
\end{align}
which leads to the Gibbs states
\begin{align}
\rho^*(\sigma,\vec{r}) &=\frac{e^{-\beta \sigma \overline{\psi}(\vec{r})-\alpha(\sigma)}}{\partfun_{\alpha}(\beta\overline{\psi}(\vec{r}))},
\intertext{where the coarse grained stream function $\bar{\psi}$ and the partition function $\partfun_{\alpha}$ are given by}
\overline{\psi}&=-\Delta^{-1}\overline{\omega}, & \partfun_{\alpha}(u)&= \int_\R e^{- \sigma u -\alpha(\sigma)} d\sigma.
\intertext{It follows that the coarse-grained vorticity field satisfies}
\overline{\omega}(\vec{r}) &= F_{\alpha}(\beta\overline{\psi}(\vec{r})), & \text{with } F_{\alpha}(u) &= - \frac{d \ln \partfun_{\alpha}(u)}{du}. \label{meanfieldeq}
\end{align}
This is a (elliptic) partial differential equation, referred to as the \emph{mean-field equation}, characterizing the most probable coarse-grained vorticity fields. Note that the equation is of the same form as the equation defining stationary states of the Euler equation: equilibrium states form a subclass of steady-states for which the function relating vorticity and stream function is fixed by the invariants of the system. 

The equilibrium states of the system can thus be obtained by solving \eqref{meanfieldeq}. In general, this is a difficult task. Analytical solutions have been obtained in the limit of a linear function $F_{\alpha}$ (the mean-field equation then reduces to a Helmholtz equation), using the method introduced by Chavanis and Sommeria~\cite{Chavanis1996a}, which consists in decomposing the vorticity field and stream function on a basis of Laplacian eigenfunctions.
Numerical methods are also available: Turkington and Whitaker have proposed an algorithm to iteratively solve the variational problem described above~\cite{Turkington1996}, while Robert and Sommeria~\cite{Robert1992} have proposed relaxation equations where the dynamics maximize the entropy production rate, thereby reaching a maximum entropy state. We shall not describe in details these methods here, nor the solutions they yield. Note, however, that in general, they correspond to large scale coherent structures, like vortices or unidirectional (e.g. zonal) flows, depending on the geometry of the domain: for instance dipole/monopole in a rectangular domain~\cite{Chavanis1996a}, dipole/unidirectional flow in a doubly periodic domain~\cite{Bouchet2009}, Fofonoff flows on a beta-plane~\cite{Naso2011}, and solid-body rotation/dipole/quadrupole/unidirectional flow on a sphere~\cite{Herbert2012a,Herbert2012b,Herbert2013b,Qi2014}.

\subsection{Non-equivalence of ensembles}\label{enseeqsection}

\subsubsection{Statistical ensembles and variational problems}

So far we have been using exclusively the microcanonical measure 
\begin{align}
\mu_{\gamma,E}^{(N)}(d\hat{\omega})&=\delta(\macro{H}_N[\hat{\omega}]-E)\prod_{k=1}^K\delta(\macro{G}_N^{(k)}[\hat{\omega}]-\gamma_k) \mu^{(N)}(d\hat{\omega}), 
\intertext{which assigns uniform probability to microstates with a given energy and vorticity distribution, and zero probability to other microstates. We could make different choices and consider the canonical measure}
\mu_{\gamma,\beta}^{(N)}(d\hat{\omega}) &= \frac{e^{-\beta \macro{H}_N[\hat{\omega}]}}{\partfun} \prod_{k=1}^{K}\delta(\macro{G}_N^{(k)}[\hat{\omega}]-\gamma_k) \mu^{(N)}(d\hat{\omega}),
\intertext{or the grand-canonical measure}
\mu_{\alpha,\beta}^{(N)}(d\hat{\omega}) &= \frac{e^{-\beta \macro{H}_N[\hat{\omega}]-\sum_{k=1}^K \alpha_k \macro{G}_N^{(k)}[\hat{\omega}]}}{\Xi} \mu^{(N)}(d\hat{\omega}),
\end{align}
and similarly in the thermodynamic limit $N \to +\infty$.
If we replace the microcanonical measure in section~\ref{meanfieldsec} by any of these two measures, we obtain \emph{mutas mutandi} a large deviation principle for the macrostates. In the thermodynamic limit, the most probable macrostates (i.e. the equilibrium states) are therefore solutions of the following variational problems:
\begin{align}
\thermo{S}(\thermo{E},\gamma) &= \max_{\rho} \{ \meanfield{S}[\rho] \suchthat \meanfield{H}[\rho]=\thermo{E}, \forall \sigma \in \R, \meanfield{D}_\sigma[\rho]=\gamma(\sigma)\}, \label{MVP}\\
\thermo{F}(\beta,\gamma) &= \max_{\rho} \{ \meanfield{S}[\rho] -\beta \meanfield{H}[\rho] \suchthat \forall \sigma \in \R, \meanfield{D}_\sigma[\rho]=\gamma(\sigma)\},\label{CVP} \\
\thermo{J}(\beta,\alpha) &= \max_{\rho} \{ \meanfield{S}[\rho] -\beta \meanfield{H}[\rho] - \int_\R d\sigma \alpha(\sigma) \meanfield{D}_\sigma[\rho]\}, \label{GVP}
\end{align}
respectively for the microcanonical measure, the canonical measure and the grand-canonical measure. The maximized functions arise as large deviation rate functions, and the constraints stem from the definition of the ensembles as conditional probabilities and from the existence of representation functions. The entropy $S(E,\gamma)$, the \emph{free energy} $F(\beta,\gamma)$ and the \emph{grand potential} $J(\beta,\alpha)$ are referred to generically as \emph{thermodynamic potentials}.

The existence of a large deviation principle for the macrostates does not depend on the particular choice of ensemble, but the most probable macrostates may depend on this choice. The task that we set out to investigate in this section is therefore how the different ensembles are related. The discussion closely follows the references~\cite{Ellis2000,Touchette2004}.

\subsubsection{Ensemble equivalence at the macrostate level}\label{macrostateequivsec}

First of all, it is clear from the structure of the variational problem and the Lagrange multiplier rule that they all have the same critical points. However, the critical points may be of different nature: a maximizer of one variational problem may be a saddle point of another variational problem for instance.
Nevertheless, it is easily seen that a solution of a variational problem with a constraint relaxed (e.g. the canonical variational problem) is always a solution of the original constrained variational problem (e.g. the microcanonical problem). We can formalize this remark by introducing the sets of equilibrium states (i.e. solutions of the variational problems):
\begin{align}
\meanfield{MC}(\thermo{E},\gamma) &= \{ \rho \suchthat \meanfield{S}[\rho]=\thermo{S}(\thermo{E},\gamma), \meanfield{H}[\rho]=\thermo{E}, \forall \sigma \in \R, \meanfield{D}_\sigma[\rho]=\gamma(\sigma)\},\\
\meanfield{C}(\beta,\gamma) &= \{ \rho \suchthat \meanfield{S}[\rho] -\beta \meanfield{H}[\rho]=\thermo{F}(\beta,\gamma), \forall \sigma \in \R, \meanfield{D}_\sigma[\rho]=\gamma(\sigma)\}, \\
\meanfield{GC}(\beta,\alpha) &= \{\rho \suchthat \meanfield{S}[\rho] -\beta \meanfield{H}[\rho] - \int_\R d\sigma \alpha(\sigma) \meanfield{D}_\sigma[\rho] = \thermo{J}(\beta,\alpha)\}.
\end{align}
As per the above remark, we always have, 
\begin{align}
&\forall \beta,\alpha, \forall \rho \in \meanfield{GC}(\beta,\alpha), \rho \in \meanfield{C}(\beta,\meanfield{D}_\sigma[\rho]) \text{ and } \rho \in \meanfield{MC}(\meanfield{H}[\rho],\meanfield{D}_\sigma[\rho]),\\
&\forall \beta, \gamma, \forall \rho \in \meanfield{C}(\beta,\gamma), \rho \in \meanfield{MC}(\meanfield{H}[\rho],\gamma).
\end{align}
In particular, 
\begin{align}
\displaystyle \bigcup_{\beta,\alpha} \meanfield{GC}(\beta,\alpha) \subset \bigcup_{\beta,\gamma} \meanfield{C}(\beta,\gamma) \subset \bigcup_{\thermo{E},\gamma} \meanfield{MC}(\thermo{E},\gamma).
\end{align}
If the converse statements hold, i.e.
\begin{align}
\forall E, \gamma, \forall \rho \in \meanfield{MC}(\thermo{E},\gamma), \exists \beta \in \R: \rho \in \meanfield{C}(\beta,\gamma), 
\intertext{or}
\forall \beta, \gamma, \forall \rho \in \meanfield{C}(\beta,\gamma), \exists \alpha: \rho \in \meanfield{GC}(\beta,\alpha), 
\end{align}
we say, respectively, that the microcanonical and canonical ensembles are \emph{equivalent at the macrostate level} or that the canonical and grand canonical ensembles are \emph{equivalent at the macrostate level}. It is straightforward to see that it is a transitive relation, in the sense that if the microcanonical ensemble is equivalent to the canonical ensemble at the macrostate level, and if the canonical ensemble and the grand-canonical ensemble are equivalent at the macrostate level, then the microcanonical and the grand-canonical ensembles are equivalent a the macrostate level. Besides, if the grand-canonical ensemble is equivalent to the microcanonical ensemble at the macrostate level, then the canonical ensemble is equivalent to both the microcanonical and the grand-canonical ensembles at the macrostate level.

If the three ensembles are equivalent at the macrostate level, we have the equalities:
\begin{align}
\displaystyle \bigcup_{\beta,\alpha} \meanfield{GC}(\beta,\alpha) = \bigcup_{\beta,\gamma} \meanfield{C}(\beta,\gamma) = \bigcup_{\thermo{E},\gamma} \meanfield{MC}(\thermo{E},\gamma).
\end{align}

\subsubsection{Ensemble equivalence at the thermodynamic level}\label{thermoequivsec}

Due to the definition of the thermodynamic potentials through the variational problems, connections exist between them as well. For the free energy for instance, we have
\begin{align*}
\thermo{F}(\beta,\gamma)&=\max_{\rho,\meanfield{N}[\rho](\vec{x})=1}\left\{ \meanfield{S}[\rho]-\beta \meanfield{H}[\rho] \suchthat \forall \sigma \in \R, \meanfield{D}_\sigma[\rho]=\gamma(\sigma) \right\},\\
&=\max_{\thermo{E}\geq 0} \Bigg( \max_{\rho, \meanfield{N}[\rho](\vec{x})=1, \meanfield{H}[\rho]=\thermo{E}} \left\{ \meanfield{S}[\rho]-\beta E \suchthat \forall \sigma \in \R, \meanfield{D}_\sigma[\rho]=\gamma(\sigma)\right\}\Bigg),\\
&=\max_{\thermo{E} \geq 0} \left( \thermo{S}(\thermo{E},\gamma) -\beta\thermo{E}\right).\\
\intertext{This exactly means that the free energy is the \emph{Legendre-Fenchel} transform of the entropy. The Legendre-Fenchel transform is a generalization of the Legendre transform to functions which need not be differentiable and convex~\cite{RockafellarBook}. 
Denoting the Legendre-Fenchel of an arbitrary function with a star (the variable with respect to which the transform is taken should be clear from the arguments of the function), we have the compact form:}
\thermo{F}(\beta,\gamma)&=\thermo{S}^{\star}(\thermo{E},\gamma).
\intertext{Similarly,}
\thermo{J}(\beta,\alpha)&=\thermo{F}^\star(\beta,\gamma).
\end{align*}
We know that the Legendre transform is an involution~\cite{ArnoldMecaBook}. This is not necessarily the case for the Legendre-Fenchel transform, because the Legendre-Fenchel transform of an arbitrary function is always a concave function, but it is true when the function is concave. In general, we only obtain the concave hull of the original function~\cite{RockafellarBook}. Hence, the free energy $F$ is always a concave function of $\beta$ and the grand-potential $J$ is always a concave function of its arguments, while $F^\star=S^{\star\star}$ is always a concave function of $E$, and is the smallest concave function satisfying $S(E,\gamma) \leq S^{\star\star}(E,\gamma)$. The equality holds if $S$ is a concave function.
Therefore, we say that the microcanonical and canonical ensemble are \emph{equivalent at the thermodynamic level} if $S=F^\star=S^{\star\star}$, or equivalently, if $S$ is a concave function of $E$. Similarly, the grand canonical and the canonical ensembles are equivalent at the thermodynamic level if $F=J^\star=F^{\star\star}$, i.e. if $F$ is a concave function of $\gamma$.

Again, we have a transitivity property: equivalence of the grand canonical and canonical ensembles on the one hand, and of the canonical and microcanonical ensembles on the other hand implies equivalence of the grand canonical and microcanonical ensembles. Besides, if the grand canonical and the microcanonical ensembles are equivalent, then the canonical ensemble is equivalent to both the grand canonical and the microcanonical ensembles. In both these cases, the entropy $S$ is a concave function of all its arguments.

\subsubsection{Equivalence and Non-equivalence of statistical ensembles}

The notions of ensemble equivalence at the macrostate level (section~\ref{macrostateequivsec}) and at the thermodynamic level (section~\ref{thermoequivsec}) are connected. Indeed, the local concavity properties of the thermodynamic potential determine the possibility to invert the relation with the Lagrange multiplier, or in other words, the possibility that the macrostates can be obtained by solving a relaxed variational problem. Following~\cite{Ellis2000}, let us examine the three possibilities in the context of the microcanonical and canonical ensembles. Let us fix $E,\gamma$, then one of the three following assertions holds:
\begin{enumerate}[label=(\roman*)]
\item \textbf{Total Ensemble Equivalence:} If $S=S^{\star\star}$ and $S$ is not locally flat, then $\meanfield{MC}(\thermo{E}, \gamma)=\meanfield{C}(\beta, \gamma)$ for $\beta=\partial \thermo{S}/\partial \thermo{E}$.
\item \textbf{Marginal Ensemble Equivalence:} If $S=S^{\star\star}$ and $S$ is locally flat, then $\meanfield{MC}(\thermo{E}, \gamma) \subsetneq \meanfield{C}(\beta, \gamma)$ for $\beta=\partial \thermo{S}/\partial \thermo{E}$.
\item \textbf{Ensemble Inequivalence:} If $S \neq S^{\star\star}$, then $\forall \beta \in \R, \meanfield{MC}(\thermo{E}, \gamma) \cap \meanfield{C}(\beta, \gamma) = \emptyset$.
\end{enumerate}

\subsection{Large deviations for the coarse-grained vorticity field}

In section~\ref{meanfieldsec}, we have considered how the probability of the outcome of a given observable (the distribution of fine-grained vorticity) behaves when the size of the system goes to infinity. We have found that it satisfies a large deviation property, which allows us to compute the most probable outcomes (see section \ref{meanfieldeqsection}). From there, we are able to deduce what the most probable coarse-grained vorticity fields are. But can we apply the same methods directly to the coarse-graining observable to compute the most probable coarse-grained vorticity fields? In other words, can we obtain a large deviation principle directly for the observable $\mathfrak{C}$? 

\begin{figure}
\includegraphics[width=\linewidth]{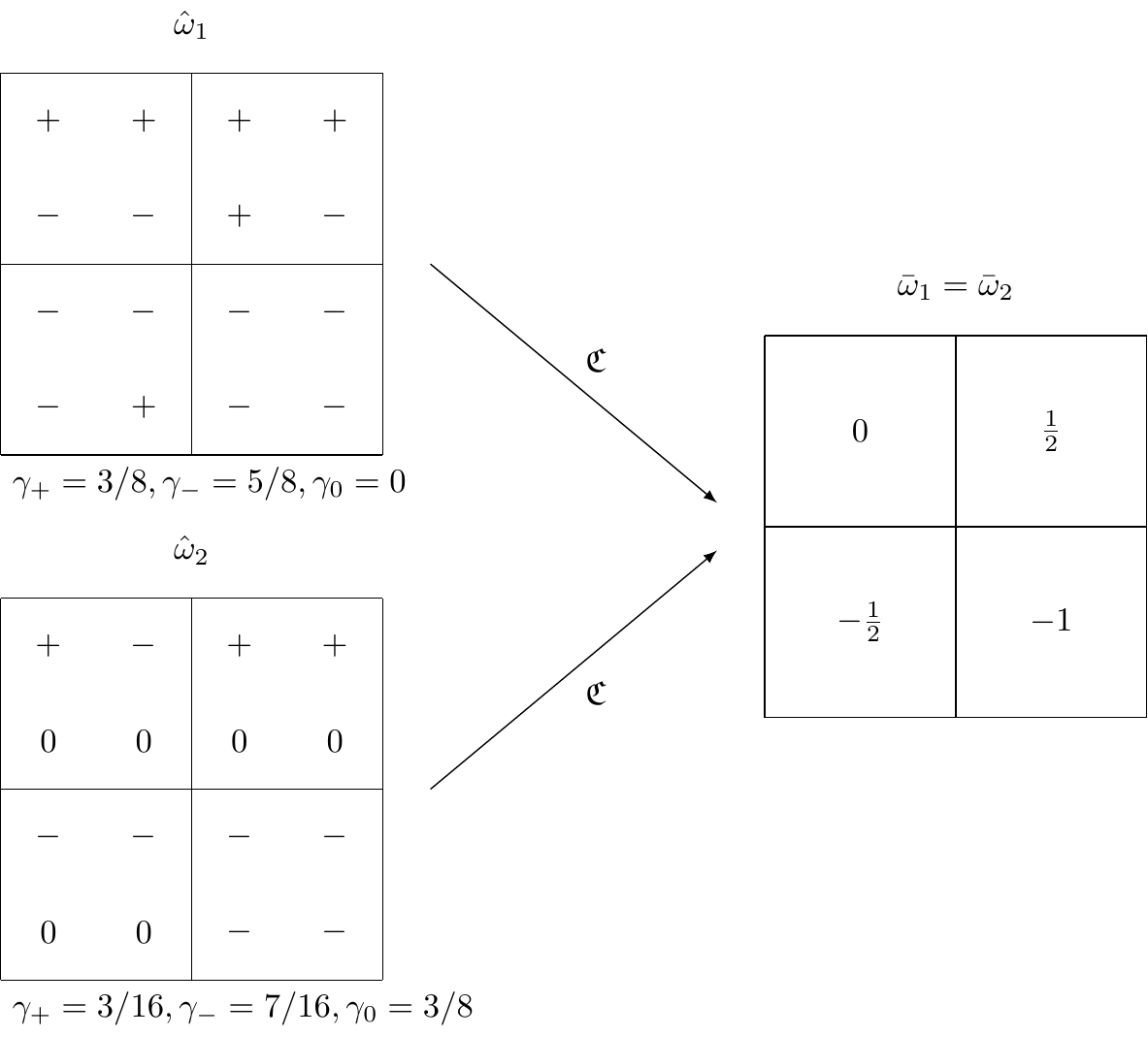}
\caption{Exemples of two microstates with different vorticity distribution, which are mapped to the same coarse-grained vorticity field by the operator $\mathfrak{C}$, in the three-level case: $\mathfrak{S}=\{-1,0,1\}$.}\label{cgfig}
\end{figure}

In general this is not straightforward, because we do not have a representation function for the vorticity distribution in terms of the coarse-grained vorticity field. Let us give an exemple in the simple case where we have only three levels of vorticity: $\mathfrak{S}=\{-1,0,1\}$. We have represented on Fig.~\ref{cgfig} two microstates which lead to the same coarse-grained vorticity field, with different vorticity distributions.
As a consequence, we cannot deduce a large deviation principle with respect to the microcanonical measure (or any of the other ensembles) from a large deviation principle with respect to the prior measure. In principle it remains possible to evaluate directly the probability of a coarse-grained vorticity field in the microcanonical ensemble, but this is a much more complicated combinatorial problem. However, in the special case of a two-level vorticity system, we do have a representation function for the vorticity distribution. We illustrate this in the following sections by making use of the analogy with the mean-field Ising model pointed out above.

\subsubsection{Mean-field Ising model}
\begin{figure}
\centering
\includegraphics[width=\linewidth]{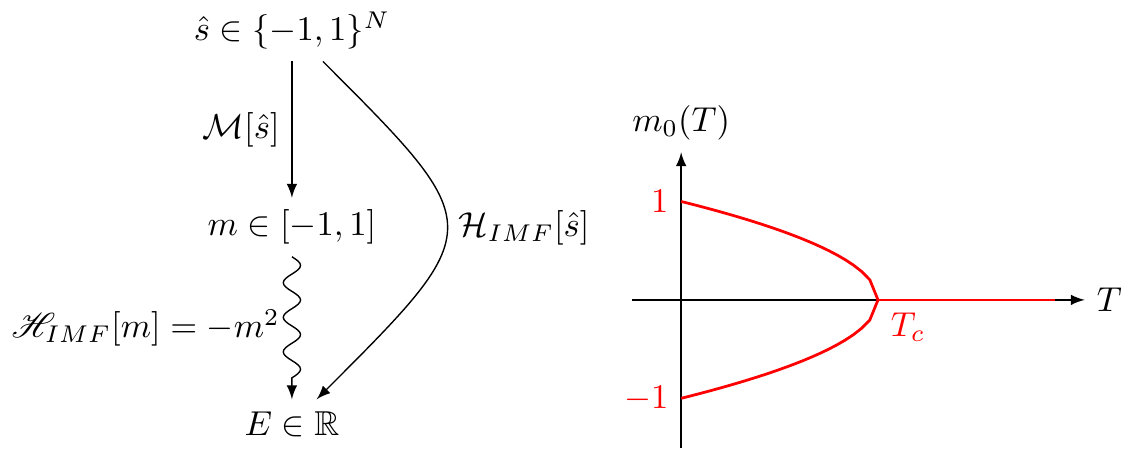}
\caption{Mean-field Ising model: observables (hamiltonian and magnetization) and representation function for the energy (left) and most probable magnetization as a function of temperature in the canonical ensemble (right).}\label{isingmffig}
\end{figure}

Remember the mean-field Ising model described in section~\ref{isingmfsec}. We have mentioned above that there is a representation function for the energy in terms of the magnetization (Fig.~\ref{isingmffig}). Therefore it is sufficient to obtain a large deviation principle for the magnetization with respect to the unconstrained measure. If $N_+$ (resp. $N_-$) is the number of $+$ (resp. $-$) spins, the magnetization is given by $\mathcal{M}[\hat{s}]=(N_+-N_-)/N$, and we have $N_++N_-=N$. In other words, $N_\pm/N=(1\pm\mathcal{M}[\hat{s}])/2$. Hence, the unconstrained probability to observe a given magnetization is
\begin{align} 
\mu^{(N)}(\mathcal{M}[\hat{s}]=m) &= \frac{N!}{2^N N_+! N_-!}\\
&= e^{N\meanfield{S}[m]+o(N)},
\intertext{where the mean-field entropy is given by (up to an unimportant constant $\ln 2$)}
\meanfield{S}[m]&= - \frac {1+m} 2 \ln \left(\frac{1+m}{2}\right) - \frac {1-m} 2 \ln \left(\frac{1-m}{2}\right),\label{isingmfentropyeq}
\end{align}
which proves that the magnetization observable satisfies a large deviation principle. It is customary to work in the canonical ensemble (see section~\ref{enseeqsection}), and the most probable states are therefore solutions of the variational problem:
\begin{align}
F(\beta) &= \min_{m \in [-1,1]} \left( \beta\meanfield{H}_{IMF}[m]-\meanfield{S}[m] \right),
\end{align}
where $F$ is the free energy. Using $\meanfield{H}_{IMF}[m]=-m^2$ and~\eqref{isingmfentropyeq}, it is easily shown that for $\beta=1/(kT)$ smaller than a critical value $\beta_c=1/(kT_c)$ (high temperature $T$), there is a unique solution $m=0$, while for $\beta$ larger than the critical value (low temperature), there are two non-zero solutions $\pm m_0(T)$. The most probable magnetization as a function of the temperature is represented on Fig.~\ref{isingmffig}.

\subsubsection{Two-level system}
We have noted above that when the vorticity level set is made of two opposite values, $\mathfrak{S}=\{\sigma_0,-\sigma_0\}$ (with $\sigma_0>0$), the system becomes analogous to the mean-field Ising model studied above. The only difference is the vorticity distribution conservation constraint (and the interaction coefficients). This amounts to keeping fixed the number of $+$ and $-$ spins in the Ising model, or equivalently, to fixing the magnetization. But the magnetization here is nothing but the circulation $\Gamma_1$. Therefore, conservation of the Casimir invariants in the discretized two-level model reduces to conservation of the circulation.

Another way to see this is to show explicitly that there exists a representation function for the vorticity distribution in this case. The coarse-graining operator $\mathfrak{C}$ takes value in a discrete subset of $\R^M$: denoting $\bar{\mathfrak{S}}_n=\{ \left(\frac{2k}{n}-1\right)\sigma_0, 0 \leq k \leq n\}$, the image of the operator is $\bar{\mathfrak{S}}_n^M$. Here, $k$ corresponds to the number of sites with value $\sigma_0$ in each coarse-graining cell. The relation between $k$ and $\bar{\omega}_i$ can be inverted: $k=n(1+\bar{\omega}_i/\sigma_0)/2$, and we obtain
\begin{align}
\gamma_+ &= \frac 1 N \sum_{i=1}^M \frac n 2 \left( 1+\frac {\bar{\omega}_i}{\sigma_0}\right),\\
&= \frac 1 2+ \frac{\Gamma_1}{2\sigma_0},
\intertext{and similarly,}
\gamma_- &= \frac 1 2 - \frac{\Gamma_1}{2\sigma_0}.
\end{align}
Note that, as expected, $\gamma_++\gamma_-=1$ (Eq.~\eqref{gammanormeq}) and $(\gamma_+-\gamma_-)\sigma_0=\Gamma_1$ (Eq.~\eqref{momvorteq}).
Now, it is an easy task to evaluate the unconstrained probability of a given coarse-grained vorticity field:
\begin{align}
\mu^{(N)}(\mathfrak{C}[\hat{\omega}]=\bar{\omega}) &= \frac {\card \mathfrak{C}^{-1}[\bar{\omega}]}{2^N} = 2^{-N} \prod_{i=1}^M \frac {n!}{(n/2(1+\bar{\omega}_i/\sigma_0))!(n/2(1-\bar{\omega}_i/\sigma_0))!},\\
&= e^{N\bar{\meanfield{S}}_{M,2}[\bar{\omega}] + o(N)},
\intertext{with the entropy of the coarse-graining observable (up to an unimportant constant $\ln 2$)}
\bar{\meanfield{S}}_{M,2}[\bar{\omega}] &= -\frac 1 {2M} \sum_{i=1}^M \left\lbrack\left( 1+\frac {\bar{\omega}_i}{\sigma_0}\right) \ln \left( 1+\frac {\bar{\omega}_i}{\sigma_0}\right) + \left( 1-\frac {\bar{\omega}_i}{\sigma_0}\right) \ln \left( 1-\frac {\bar{\omega}_i}{\sigma_0}\right) \right\rbrack.
\end{align}
The contraction principle ensures that, as can be checked explicitly,
\begin{align}
\bar{\meanfield{S}}_{M,2}[\bar{\omega}] &= \max_{P \in \mathcal{M}_{M,2}(\R)} \{ \meanfield{S}_{M,2}[P] \suchthat \mathfrak{E}[P]=\bar{\omega} \}.
\end{align}
By the same token as in section~\ref{meanfieldsec}, it follows that the most probable coarse-grained vorticity fields are solutions of the constrained variational problem:
\begin{align}
S(E,\gamma) &= \max_{\bar{\omega}\in \bar{\mathfrak{S}}_n^M} \{ \bar{\meanfield{S}}_{M,2}[\bar{\omega}] \suchthat \bar{\macro{H}}_M[\bar{\omega}]=E, \bar{\macro{G}}_M^{(+)}[\bar{\omega}]=\gamma_+, \bar{\macro{G}}_M^{(-)}[\bar{\omega}]=\gamma_-\},
\intertext{or equivalently,}
S(E,\Gamma_1) &= \max_{\bar{\omega}\in \bar{\mathfrak{S}}_n^M} \{ \bar{\meanfield{S}}_{M,2}[\bar{\omega}] \suchthat \bar{\macro{H}}_M[\bar{\omega}]=E, \bar{\macro{M}}_M[\bar{\omega}]=\Gamma_1\}.
\end{align}
Straightforward computations show that the critical points of the variational problem are solutions of the equation:
\begin{align}
\bar{\omega} &= \sigma_0 \tanh \left( \frac{(\beta \bar{\psi}+\alpha_1)\sigma_0}{2}\right).
\end{align}

\subsubsection{Fragile constraints and constrained Casimir variational problem}

It has been observed by several authors that none of the Casimir invariants $\Gamma_p$ (moments of the vorticity field) except the first (circulation) can be obtained from the coarse-grained vorticity field $\bar{\omega}$. For this reason they are often referred to as \emph{fragile invariants}, in the sense that they do not survive coarse-graining.
 This is exactly the same as saying that there is no representation function for the Casimir invariants (or, equivalently, for the vorticity distribution $\gamma(\sigma)$) in terms of the coarse-grained vorticity field $\bar{\omega}$, except in the particular case mentioned above. However, with an arbitrary vorticity distribution, a large deviation principle can still be obtained by contraction, as illustrated above in the two-layer case. This provides a variational problem for the most probable coarse-grained vorticity field, even though it still relies on an auxiliary maximization on the distribution $P$ for the vorticity levels.

Because of their fragile nature, and because an infinite number of invariants is difficult to handle in practice, it was suggested~\cite{Ellis2002,Chavanis2003} to treat these invariants in a canonical ensemble, and to consider the Lagrange parameter $\alpha(\sigma)$ as a prior vorticity distribution chosen based on physical intuition of the problem at hand. This provides a subset of solutions of the microcanonical variational problem, but not necessarily the full set (see section~\ref{enseeqsection}). However, this variational problem, expressed in terms of the distribution $P$ for the vorticity levels, is equivalent to minimizing, with respect to the coarse-grained vorticity field $\bar{\omega}$, the so-called \emph{Casimir functionals} $\int_\domain s(\omega)$ with fixed energy, where $s$ is a convex function, choosing for $s$ the Legendre-Fenchel transform of $\ln \partfun_\alpha$~\cite{Bouchet2008}.

\subsection{The energy-enstrophy measure}\label{energyenstrophysection}

\subsubsection{Gibbs measure for Galerkin truncated flows}\label{kraichnansection}

In this section we investigate the statistical mechanics of the 2D Euler equations resulting from simplifying the conservation constraints: we retain only the energy and the enstrophy invariants. This was actually one of the starting points for statistical mechanics of turbulent flows: Lee in 3D~\cite{Lee1952} and Kraichnan in 2D~\cite{Kraichnan1967} considered Fourier series of the dynamical fields truncated at a given order $N$. In 3D, the only invariants are the energy and the helicity, while in 2D, Kraichnan considered the energy:
\begin{align}
\mathcal{H}_N[\hat{\omega}] &= \frac 1 2 \sum_{i=1}^N \frac{\abs{\hat{\omega}_i}^2}{\lambda_i},
\intertext{and the enstrophy}
\mathcal{G}_N[\hat{\omega}] &= \frac 1 2 \sum_{i=1}^N \abs{\hat{\omega}_i}^2,
\intertext{where the truncated vorticity field is given by}
\hat{\omega}(x) &= \sum_{i=1}^N \hat{\omega}_i \phi_i(x), \quad \Delta \phi_i = -\lambda_i \phi_i, 0\leq \lambda_i \leq \lambda_{i+1}.
\intertext{It is assumed that the truncated vorticity field $\hat{\omega}$ is a random variable distributed according to the canonical (Gibbs) measure:}
\mu_{\beta,\alpha}^{(N)}(d\hat{\omega}) &= \frac 1 {\partfun} e^{-\beta \mathcal{H}_N[\hat{\omega}] - \alpha \mathcal{G}_N[\hat{\omega}]} \prod_{i=1}^Nd\hat{\omega}_i.
\end{align}
This is a Gaussian probability density, well-defined if $\beta+\alpha \lambda_i >0$ for all $i$. This condition leads to three possible regimes: (i) $\beta<0, \alpha >0$, (ii) $\beta>0, \alpha>0$, (iii) $\beta>0, \alpha<0$. In each case, Kraichnan considered the energy spectrum $\macro{E}_i[\hat{\omega}]=\frac{\abs{\hat{\omega}_i}^2}{2\lambda_i}$ and computed its average value with respect to the Gibbs measure~\cite{Kraichnan1967,Kraichnan1980}: $\langle \macro{E}_i \rangle = \frac{1}{2(\beta+\alpha\lambda_i)}$. In the negative temperature ($\beta<0$) regime, the spectrum peaks at the gravest mode $\phi_1$; there is even an \emph{infrared divergence} when $\beta \to -\alpha \lambda_1$. This is classically interpreted as an indication that not only nonlinear interactions in 2D flows tend to transfer energy towards the large scales (the \emph{inverse cascade}), but there is a tendency for energy to accumulate in the gravest mode to form a \emph{condensate}~\cite{LSmith1993,Chertkov2007,Boffetta2012}.
Note that the average value of each vorticity mode vanishes by symmetry: $\langle \hat{\omega}_i \rangle=0$, because a given vorticity field and its opposite have the same probability in the canonical ensemble. Of course, in reality, the system will spontaneously break the symmetry and choose a vorticity field, which can be computed in the limit of large $N$ using large deviations results for the macrostates as we did above. In the energy-enstrophy ensemble, averaging over the set of equilibrium states indeed yields a vanishing mean value, thereby showing that statistical mechanics is more about most probable states than average values.

\subsubsection{Large deviations in the microcanonical ensemble}

Using the same notations as in the previous paragraph, one may assume that the truncated vorticity field is distributed according to the microcanonical measure 
\begin{align}
\mu_{E,\Gamma_2}^{(N)}(d\hat{\omega}) &= \frac 1 {\Omega_N(E,\Gamma_2)} \delta(\mathcal{H}_N[\hat{\omega}]-E) \delta(\mathcal{G}_N[\hat{\omega}]-\Gamma_2) \prod_{i=1}^Nd\hat{\omega}_i,
\intertext{instead of the Gibbs measure $\mu_{\beta,\alpha}^{(N)}(d\hat{\omega})$, with the structure function given by}
\Omega_N(E,\Gamma_2) &= \int \delta(\mathcal{H}_N[\hat{\omega}]-E) \delta(\mathcal{G}_N[\hat{\omega}]-\Gamma_2) \prod_{i=1}^Nd\hat{\omega}_i.
\intertext{Rather exceptionally, since both constraints involve quadratic functions, the structure function can be computed explicitly, using integral representations of the Dirac distributions~\cite{Bouchet2010}, and thus also the entropy:}
\Omega_N(E,\Gamma_2) &= e^{N S(E,\Gamma_2)+o(N)},\\
S(E,\Gamma_2) &= \frac 1 2 \ln (\Gamma_2-2\lambda_1 E). \label{energyenstrophyentropyeq}
\end{align}
Note that Bouchet and Corvellec~\cite{Bouchet2010} have also checked with explicit computations that this entropy defined as the joint large deviation rate function for the energy and enstrophy observables (i.e. the Boltzmann formula $S(E,\Gamma_2)=\lim (1/N) \ln \Omega_N(E,\Gamma_2)$, given in \eqref{energyenstrophyentropyeq}) coincides with the entropy defined through the variational problem for the macrostates, as expected from the contraction principle (Eq.~\eqref{mcvpeq}).
A similar computation of the structure function was carried out by Kastner and Schnetz~\cite{Kastner2006} for the mean-field spherical model defined in section~\ref{sphericalmodelsec}.

From the joint large deviation principle for the energy and enstrophy, we can deduce a large deviation principle for the energy spectrum observable~\cite{Bouchet2010}:
\begin{align}
\mu^{(N)}_{E,\Gamma_2}(\macro{E}_i[\hat{\omega}]=E_i)&=e^{N\meanfield{S}^{(i)}_{E,\Gamma_2}[E_i]+o(N)},\\
\intertext{with}
\meanfield{S}^{(1)}_{E,\Gamma_2}[E_1] &=
\begin{cases}
\frac 1 2 \ln(\Gamma_2-2\lambda_2 E+2(\lambda_2-\lambda_1)E_1) & \text{ if } 0<E_1<E\\
-\infty												 & \text{ otherwise}
\end{cases},
\intertext{and for $i>1$,}
\meanfield{S}^{(i)}_{E,\Gamma_2}[E_i] &=
\begin{cases}
\frac 1 2 \ln(\Gamma_2-2\lambda_1 E-2(\lambda_i-\lambda_1)E_i) & \text{ if } 0 \leq E_i \leq E\\
-\infty												 & \text{ otherwise}
\end{cases}.
\end{align}
The large deviation rate functions are monotonous: $\meanfield{S}^{(1)}_{E,\Gamma_2}[E_1]$ is an increasing function of $E_1$, and $\meanfield{S}^{(i)}_{E,\Gamma_2}[E_i]$ are decreasing functions of $E_i$. Therefore, the most probable energy spectrum in the limit $N\to +\infty$ has all its energy in the gravest mode. This can be seen as the microcanonical counterpart of the Kraichnan argument presented in section~\ref{kraichnansection}. It provides further theoretical evidence for the spectral condensation in 2D turbulence.

The above discussion on the vanishing of the average truncated vorticity field also applies in the microcanonical ensemble. The mean-field theory allows to compute the most probable macrostates: we find a linear mean-field equation for the coarse-grained vorticity field: $\bar{\omega}=\beta/(2\alpha) \bar{\psi}$, which is easily solved and yields $\bar{\omega}=\sqrt{2\lambda_1 E}\phi_1$, in agreement with the above prediction.

\section{Conclusion}

In this chapter, we have given a brief introduction to the methods of equilibrium statistical mechanics applied to models of turbulent flows, focusing on the case of two-dimensional flows. The main purpose of the course was to show, in the context of a lattice discretization of the system, how some well-chosen observables, such as the distribution of fined-grained vorticity levels, concentrate in probability around a set of equilibrium values. Such properties are conveniently expressed using the theory of large deviations. In fact, we have closely followed the principles of equilibrium statistical mechanics formulated in the language of large deviations, as exposed for instance in~\cite{Touchette2009}. A major ingredient in deriving the large deviation results is the long-range character of the interactions, because it leads to the existence of a representation function for the energy. This is a major simplification, as it allows us to compute the probability of a macrostate with respect to the uniform measure and then deduce the probability with respect to the microcanonical measure. 
We have emphasized this point by considering another observable, the coarse-grained vorticity field (for which there is no representation function for the vorticity distribution) and by making the analogy with a simpler system, the mean-field Ising model.

The large deviation principle leads to a variational problem characterizing the most probable macrostates. This allows to compute coarse-grained vorticity fields which should correspond in practice to the final state of the system, if ergodicity holds. This provides a statistical explanation of the spontaneous emergence of coherent structures in two-dimensional flows. The equilibrium states obtained may depend on the choice of probability measure in phase space: we have discussed the relations between the standard ensembles of statistical mechanics and given a connection with the concavity properties of the entropy.

In the simpler context of the energy-enstrophy measure, we have explained that the energy spectrum observable also satisfies a large deviation principle, which shows that the most probable state has all its energy condensed in the gravest mode. This is physically consistent with the familiar ideas of inverse cascade of energy and energy condensation for 2D flows.

\bibliographystyle{spmpsci}
\bibliography{bibtexlib}

\begin{thebibliography}{10}
\providecommand{\url}[1]{{#1}}
\providecommand{\urlprefix}{URL }
\expandafter\ifx\csname urlstyle\endcsname\relax
  \providecommand{\doi}[1]{DOI~\discretionary{}{}{}#1}\else
  \providecommand{\doi}{DOI~\discretionary{}{}{}\begingroup
  \urlstyle{rm}\Url}\fi

\bibitem{ArnoldMecaBook}
Arnold, V.I.: {Mathematical Methods of Classical Mechanics}, 2nd edition edn.
\newblock Springer (1989)

\bibitem{Batchelor1969}
Batchelor, G.: Computation of the energy spectrum in homogeneous
  two-dimensional turbulence.
\newblock Phys. Fluids \textbf{12}(Suppl. II), 233--239 (1969).
\newblock \doi{10.1063/1.1692443}

\bibitem{BaxterBook}
Baxter, R.J.: {Exactly Solved Models in Statistical Mechanics}.
\newblock Academic Press (1982)

\bibitem{Berlin1952}
Berlin, T.H., Kac, M.: {The spherical model of a ferromagnet}.
\newblock Phys. Rev. \textbf{86}, 821 (1952).
\newblock \doi{10.1103/PhysRev.86.821}

\bibitem{Biferale2012}
Biferale, L., Musacchio, S., Toschi, F.: {Inverse Energy Cascade in
  Three-Dimensional Isotropic Turbulence}.
\newblock Phys. Rev. Lett. \textbf{108}(16), 164,501 (2012).
\newblock \doi{10.1103/PhysRevLett.108.164501}.
\newblock
  \urlprefix\url{http://link.aps.org/doi/10.1103/PhysRevLett.108.164501}

\bibitem{Boffetta2012}
Boffetta, G., Ecke, R.E.: {Two-Dimensional Turbulence}.
\newblock Ann. Rev. Fluid Mech. \textbf{44}, 427 (2012).
\newblock \doi{10.1146/annurev-fluid-120710-101240}

\bibitem{Boucher1999}
Boucher, C., Ellis, R.S., Turkington, B.: Spatializing random measures: Doubly
  indexed processes and the large deviation principle.
\newblock Annals of Probability \textbf{27}, 297--324 (1999)

\bibitem{Boucher2000}
Boucher, C., Ellis, R.S., Turkington, B.: Derivation of maximum entropy
  principles in two-dimensional turbulence via large deviations.
\newblock J. Stat. Phys. \textbf{98}(5-6), 1235--1278 (2000)

\bibitem{Bouchet2008}
Bouchet, F.: {Simpler variational problems for statistical equilibria of the 2D
  Euler equation and other systems with long range interactions}.
\newblock Physica D \textbf{237}, 1976--1981 (2008).
\newblock \doi{10.1016/j.physd.2008.02.029}

\bibitem{Bouchet2010}
Bouchet, F., Corvellec, M.: {Invariant measures of the 2D Euler and Vlasov
  equations}.
\newblock J. Stat. Mech. \textbf{2010}, P08,021 (2010).
\newblock \doi{10.1088/1742-5468/2010/08/P08021}

\bibitem{Bouchet2009}
Bouchet, F., Simonnet, E.: Random changes of flow topology in two-dimensional
  and geophysical turbulence.
\newblock Phys. Rev. Lett. \textbf{102}, 94,504 (2009).
\newblock \doi{10.1103/PhysRevLett.102.094504}

\bibitem{Bouchet2002}
Bouchet, F., Sommeria, J.: Emergence of intense jets and {J}upiter's {G}reat
  {R}ed {S}pot as maximum-entropy structures.
\newblock J. Fluid Mech. \textbf{464}, 165--207 (2002).
\newblock \doi{10.1017/S0022112002008789}

\bibitem{Bouchet2012}
Bouchet, F., Venaille, A.: Statistical mechanics of two-dimensional and
  geophysical flows.
\newblock Phys. Rep. \textbf{515}, 227--295 (2012).
\newblock \doi{10.1016/j.physrep.2012.02.001}

\bibitem{Campa2009}
Campa, A., Dauxois, T., Ruffo, S.: Statistical mechanics and dynamics of
  solvable models with long-range interactions.
\newblock Phys. Rep. \textbf{480}, 57--159 (2009).
\newblock \doi{10.1016/j.physrep.2009.07.001}

\bibitem{Charney1971}
Charney, J.G.: {Geostrophic Turbulence}.
\newblock J. Atmos. Sci. \textbf{28}, 1087--1094 (1971).
\newblock \doi{10.1175/1520-0469(1971)028<1087:GT>2.0.CO;2}

\bibitem{Chavanis2003}
Chavanis, P.: Generalized thermodynamics and fokker-planck equations:
  Applications to stellar dynamics and two-dimensional turbulence.
\newblock Phys. Rev. E \textbf{68}, 36,108 (2003)

\bibitem{Chavanis2006h}
Chavanis, P.H.: {Phase transitions in self-gravitating systems}.
\newblock Int. J. Mod. Phys. B \textbf{20}, 3113 (2006).
\newblock \doi{10.1142/S0217979206035400}

\bibitem{Chavanis2006b}
Chavanis, P.H., Dubrulle, B.: Statistical mechanics of the shallow-water system
  with an a priori potential vorticity distribution.
\newblock C. R. Phys. \textbf{7}, 422--432 (2006).
\newblock \doi{10.1016/j.crhy.2006.01.007}

\bibitem{Chavanis1996a}
Chavanis, P.H., Sommeria, J.: Classification of self-organized vortices in
  two-dimensional turbulence: the case of a bounded domain.
\newblock J. Fluid Mech. \textbf{314}, 267--297 (1996).
\newblock \doi{10.1017/S0022112096000316}

\bibitem{Chavanis2002a}
Chavanis, P.H., Sommeria, J.: Statistical mechanics of the shallow water
  system.
\newblock Phys. Rev. E \textbf{65}, 026,302 (2002).
\newblock \doi{10.1103/PhysRevE.65.026302}

\bibitem{Chertkov2007}
Chertkov, M., Connaughton, C., Kolokolov, I., Lebedev, V.: {Dynamics of Energy
  Condensation in Two-Dimensional Turbulence}.
\newblock Phys. Rev. Lett. \textbf{99}(8), 084,501 (2007).
\newblock \doi{10.1103/PhysRevLett.99.084501}.
\newblock \urlprefix\url{http://link.aps.org/doi/10.1103/PhysRevLett.99.084501}

\bibitem{Cipra1987}
Cipra, B.A.: {An introduction to the Ising model}.
\newblock Amer. Math. Monthly \textbf{94}(10), 937--959 (1987)

\bibitem{DauxoisLRIbook}
Dauxois, T., Ruffo, S., Arimondo, E., Wilkens, M. (eds.): Dynamics and
  Thermodynamics of Systems with Long Range Interactions, \emph{Lecture Notes
  in Physics}, vol. 602.
\newblock Springer, New-York (2002).
\newblock \doi{10.1007/3-540-45835-2}

\bibitem{DiBattista2001a}
DiBattista, M., Majda, A.J.: Equilibrium statistical predictions for baroclinic
  vortices: The role of angular momentum.
\newblock Theor. Comput. Fluid Dyn. \textbf{14}(5), 293--322 (2001)

\bibitem{EllisBook}
Ellis, R.S.: Entropy, Large Deviations, and Statistical Mechanics.
\newblock Springer, New-York (1985)

\bibitem{Ellis2000}
Ellis, R.S., Haven, K., Turkington, B.: Large deviation principles and complete
  equivalence and nonequivalence results for pure and mixed ensembles.
\newblock J. Stat. Phys. \textbf{101}, 999--1064 (2000).
\newblock \doi{10.1023/A:1026446225804}

\bibitem{Ellis2002}
Ellis, R.S., Haven, K., Turkington, B.: Nonequivalent statistical equilibrium
  ensembles and refined stability theorems for most probable flows.
\newblock Nonlinearity \textbf{15}, 239 (2002).
\newblock \doi{10.1088/0951-7715/15/2/302}

\bibitem{Eyink2006}
Eyink, G., Sreenivasan, K.: Onsager and the theory of hydrodynamic turbulence.
\newblock Rev. Mod. Phys. \textbf{78}, 87--135 (2006).
\newblock \doi{10.1103/RevModPhys.78.87}

\bibitem{FalkovichBook}
Falkovich, G.: {Fluid Mechanics}.
\newblock Cambridge University Press (2011)

\bibitem{Fjortoft1953}
Fjortoft, R.: {On the changes in the spectral distribution of kinetic energy
  for twodimensional, nondivergent flow}.
\newblock Tellus \textbf{5}, 225--230 (1953)

\bibitem{FrischBook}
Frisch, U.: Turbulence, the legacy of A.N. Kolmogorov.
\newblock Cambridge University Press (1995)

\bibitem{Herbert2013b}
Herbert, C.: {Additional invariants and statistical equilibria for the 2D Euler
  equations on a spherical domain}.
\newblock J. Stat. Phys. \textbf{152}, 1084--1114 (2013).
\newblock \doi{10.1007/s10955-013-0809-6}

\bibitem{Herbert2014b}
Herbert, C.: {Nonlinear energy transfers and phase diagrams for geostrophically
  balanced rotating-stratified flows}.
\newblock Phys. Rev. E \textbf{89}, 033,008 (2014).
\newblock \doi{10.1103/PhysRevE.89.033008}

\bibitem{Herbert2014a}
Herbert, C.: {Restricted Partition Functions and Inverse Energy Cascades in
  parity symmetry breaking flows}.
\newblock Phys. Rev. E \textbf{89}, 013,010 (2014).
\newblock \doi{10.1103/PhysRevE.89.013010}

\bibitem{Herbert2012a}
Herbert, C., Dubrulle, B., Chavanis, P.H., Paillard, D.: Phase transitions and
  marginal ensemble equivalence for freely evolving flows on a rotating sphere.
\newblock Phys. Rev. E \textbf{85}, 056,304 (2012).
\newblock \doi{10.1103/PhysRevE.85.056304}

\bibitem{Herbert2012b}
Herbert, C., Dubrulle, B., Chavanis, P.H., Paillard, D.: Statistical mechanics
  of quasi-geostrophic flows on a rotating sphere.
\newblock J. Stat. Mech. \textbf{2012}, P05,023 (2012).
\newblock \doi{10.1088/1742-5468/2012/05/P05023}

\bibitem{Herbert2014c}
Herbert, C., Pouquet, A., Marino, R.: {Restricted Equilibrium and the Energy
  Cascade in Rotating and Stratified Flows}.
\newblock J. Fluid Mech. \textbf{758}, 374--406 (2014).
\newblock \doi{10.1017/jfm.2014.540}

\bibitem{Ising1925}
Ising, E.: {Beitrag zur theorie des ferromagnetismus}.
\newblock Z. Physik. \textbf{31}, 253--258 (1925)

\bibitem{Joyce1966}
Joyce, G.S.: {Spherical model with long-range ferromagnetic interactions}.
\newblock Phys. Rev. \textbf{146}, 349 (1966)

\bibitem{Kastner2006}
Kastner, M., Schnetz, O.: On the mean-field spherical model.
\newblock J. Stat. Phys. \textbf{122}, 1195--1214 (2006)

\bibitem{Kolmogorov1941a}
Kolmogorov, A.N.: {The local structure of turbulence in incompressible viscous
  fluid for very large Reynolds' numbers}.
\newblock Dokl. Akad. Nauk. SSSR \textbf{30}, 301 (1941).
\newblock \doi{10.1098/rspa.1991.0075}

\bibitem{Kraichnan1967}
Kraichnan, R.H.: Inertial ranges in two-dimensional turbulence.
\newblock Phys. Fluids \textbf{10}, 1417--1423 (1967).
\newblock \doi{10.1063/1.1762301}

\bibitem{Kraichnan1973}
Kraichnan, R.H.: {Helical turbulence and absolute equilibrium}.
\newblock J. Fluid Mech. \textbf{59}, 745--752 (1973)

\bibitem{Kraichnan1980}
Kraichnan, R.H., Montgomery, D.C.: Two-dimensional turbulence.
\newblock Rep. Prog. Phys. \textbf{43}, 547 (1980).
\newblock \doi{10.1088/0034-4885/43/5/001}

\bibitem{LandauFluidBook}
Landau, L., Lifchitz, E.: {Physique Th\'eorique, Tome VI: M\'ecanique des
  fluides}.
\newblock Mir, Moscou (1971)

\bibitem{LanfordBook}
Lanford, O.E.: {Entropy and equilibrium states in classical statistical
  mechanics}.
\newblock In: A.~Lenard (ed.) {Statistical Mechanics and Mathematical
  Problems}, \emph{Lecture Notes in Physics}, vol.~20, pp. 1--113. Springer,
  Berlin (1973)

\bibitem{Lee1952}
Lee, T.D.: On some statistical properties of hydrodynamical and
  magneto-hydrodynamical fields.
\newblock Q. Appl. Math. \textbf{10}, 69--74 (1952)

\bibitem{Leith1968}
Leith, C.: {Diffusion Approximation for Two-Dimensional Turbulence}.
\newblock Phys. Fluids \textbf{11}, 671--673 (1968).
\newblock \doi{10.1063/1.1691968}

\bibitem{Leprovost2006}
Leprovost, N., Dubrulle, B., Chavanis, P.H.: Dynamics and thermodynamics of
  axisymmetric flows: Theory.
\newblock Phys. Rev. E \textbf{73}, 46,308 (2006).
\newblock \doi{10.1103/PhysRevE.73.046308}

\bibitem{Lilly1983}
Lilly, D.K.: {Stratified turbulence and the mesoscale variability of the
  atmosphere.}
\newblock J. Atmos. Sci. \textbf{40}, 749--761 (1983)

\bibitem{Lim2001c}
Lim, C.C.: {A long range spherical model and exact solutions of an energy
  enstrophy theory for two-dimensional turbulence}.
\newblock Physics of Fluids \textbf{13}, 1961 (2001)

\bibitem{Lim2012}
Lim, C.C.: {Phase transition to super-rotating atmospheres in a simple
  planetary model for a nonrotating massive planet: Exact solution}.
\newblock Phys. Rev. E \textbf{86}(6), 066,304 (2012).
\newblock \doi{10.1103/PhysRevE.86.066304}.
\newblock \urlprefix\url{http://link.aps.org/doi/10.1103/PhysRevE.86.066304}

\bibitem{Lucarini2014}
Lucarini, V., Blender, R., Herbert, C., Pascale, S., Ragone, F., Wouters, J.:
  {Mathematical and Physical Ideas for Climate Science}.
\newblock Rev. Geophys. \textbf{52}, 809--859 (2014).
\newblock \doi{10.1002/2013RG000446}

\bibitem{MajdaWangBook}
Majda, A.J., Wang, X.: {Nonlinear Dynamics and Statistical Theories for Basic
  Geophysical Flows}.
\newblock Cambridge University Press, Cambridge (2006)

\bibitem{McWilliams1984}
McWilliams, J.C.: The emergence of isolated coherent vortices in turbulent
  flow.
\newblock J. Fluid Mech. \textbf{146}, 21--43 (1984).
\newblock \doi{10.1017/S0022112084001750}

\bibitem{Merilees1975}
Merilees, P.E., Warn, H.: {On energy and enstrophy exchanges in two-dimensional
  non-divergent flow}.
\newblock J. Fluid Mech. \textbf{69}(04), 625--630 (1975)

\bibitem{Michel1994b}
Michel, J., Robert, R.: Large deviations for {Y}oung measures and statistical
  mechanics of infinite dimensional dynamical systems with conservation law.
\newblock Commun. Math. Phys. \textbf{159}, 195--215 (1994).
\newblock \doi{10.1007/BF02100491}

\bibitem{Michel1994a}
Michel, J., Robert, R.: Statistical mechanical theory of the great red spot of
  {J}upiter.
\newblock J. Stat. Phys. \textbf{77}, 645--666 (1994).
\newblock \doi{10.1007/BF02179454}

\bibitem{Miller1990}
Miller, J.: Statistical mechanics of {E}uler equations in two dimensions.
\newblock Phys. Rev. Lett. \textbf{65}, 2137--2140 (1990).
\newblock \doi{10.1103/PhysRevLett.65.2137}

\bibitem{Miller1992}
Miller, J., Weichman, P.B., Cross, M.C.: Statistical mechanics, {E}uler's
  equation, and {J}upiter's {R}ed {S}pot.
\newblock Phys. Rev. A \textbf{45}, 2328--2359 (1992).
\newblock \doi{10.1103/PhysRevA.45.2328}

\bibitem{Naso2011}
Naso, A., Chavanis, P.H., Dubrulle, B.: {Statistical mechanics of Fofonoff
  flows in an oceanic basin}.
\newblock Eur. Phys. J. B \textbf{80}, 493--517 (2011).
\newblock \doi{10.1140/epjb/e2011-10440-8}

\bibitem{Naso2010b}
Naso, A., Monchaux, R., Chavanis, P.H., Dubrulle, B.: {Statistical mechanics of
  Beltrami flows in axisymmetric geometry: Theory reexamined}.
\newblock Phys. Rev. E \textbf{81}, 066,318 (2010).
\newblock \doi{10.1103/PhysRevE.81.066318}

\bibitem{NazarenkoBook}
Nazarenko, S.V.: {Wave Turbulence}, \emph{Lecture Notes in Physics}, vol. 825.
\newblock Springer (2010)

\bibitem{OlverBook}
Olver, P.J.: Applications of Lie Groups to Differential Equations.
\newblock Graduate Texts in Mathematics. Springer (2000)

\bibitem{Onsager1944}
Onsager, L.: {Crystal statistics. I. A two-dimensional model with an
  order-disorder transition}.
\newblock Physical Review \textbf{65}, 117 (1944)

\bibitem{Padmanabhan1990}
Padmanabhan, T.: {Statistical mechanics of gravitating systems}.
\newblock Phys. Rep. \textbf{188}, 285--362 (1990).
\newblock \doi{10.1016/0370-1573(90)90051-3}

\bibitem{Paret1997}
Paret, J., Tabeling, P.: {Experimental observation of the two-dimensional
  inverse energy cascade}.
\newblock Phys. Rev. Lett. \textbf{79}, 4162--4165 (1997)

\bibitem{Penrose1979}
Penrose, O., Lebowitz, J.L.: Towards a rigorous molecular theory of
  metastability.
\newblock In: E.W. Montroll, J.L. Lebowitz (eds.) Fluctuation Phenomena,
  chap.~5, p. 293. Amsterdam: North-Holland (1979)

\bibitem{Potters2013}
Potters, M., Vaillant, T., Bouchet, F.: {Sampling microcanonical measures of
  the 2D Euler equations through Creutz's algorithm: a phase transition from
  disorder to order when energy is increased}.
\newblock J. Stat. Mech. (02), P02,017 (2013).
\newblock \doi{10.1088/1742-5468/2013/02/P02017}.
\newblock
  \urlprefix\url{http://stacks.iop.org/1742-5468/2013/i=02/a=P02017?key=crossref.813285bf80ff263d3ceb4b3fd761a4a8}

\bibitem{Pouquet2013}
Pouquet, A., Marino, R.: {Geophysical turbulence and the duality of the energy
  flow across scales}.
\newblock Phys. Rev. Lett. \textbf{111}, 234,501 (2013).
\newblock \doi{10.1103/PhysRevLett.111.234501}

\bibitem{Qi2014}
Qi, W., Marston, J.B.: {Hyperviscosity and statistical equilibria of Euler
  turbulence on the torus and the sphere}.
\newblock J. Stat. Mech. p. P07020 (2014).
\newblock \doi{10.1088/1742-5468/2014/07/P07020}.
\newblock \urlprefix\url{http://arxiv.org/abs/1312.2553v1}

\bibitem{Rhines1979}
Rhines, P.B.: {Geostrophic turbulence}.
\newblock Ann. Rev. Fluid Mech. \textbf{11}, 401--441 (1979)

\bibitem{Robert1989}
Robert, R.: {Concentration et entropie pour les mesures d'Young}.
\newblock C.R. Acad. Sci. Paris, S\'erie I \textbf{309}, 757 (1989)

\bibitem{Robert1990}
Robert, R.: Etats d'\'equilibre statistique pour l'\'ecoulement bidimensionnel
  d'un fluide parfait.
\newblock C.R. Acad. Sci. Paris, S\'erie I \textbf{311}, 575 (1990)

\bibitem{Robert1991b}
Robert, R.: A maximum-entropy principle for two-dimensional perfect fluid
  dynamics.
\newblock J. Stat. Phys. \textbf{65}, 531--553 (1991).
\newblock \doi{10.1007/BF01053743}

\bibitem{Robert2000}
Robert, R.: On the statistical mechanics of 2{D} {E}uler equation.
\newblock Commun. Math. Phys. \textbf{212}, 245--256 (2000).
\newblock \doi{10.1007/s002200000210}

\bibitem{Robert1991a}
Robert, R., Sommeria, J.: Statistical equilibrium states for two-dimensional
  flows.
\newblock J. Fluid Mech. \textbf{229}, 291--310 (1991).
\newblock \doi{10.1017/S0022112091003038}

\bibitem{Robert1992}
Robert, R., Sommeria, J.: Relaxation towards a statistical equilibrium state in
  two-dimensional perfect fluid dynamics.
\newblock Phys. Rev. Lett. \textbf{69}, 2776--2779 (1992).
\newblock \doi{10.1103/PhysRevLett.69.2776}

\bibitem{RockafellarBook}
Rockafellar, R.: {Convex Analysis}.
\newblock Princeton University Press (1970)

\bibitem{RuelleBook}
Ruelle, D.: Statistical Mechanics: Rigorous Results.
\newblock Amsterdam: Benjamin (1969)

\bibitem{RuelleBook1989}
Ruelle, D.: {Chaotic Evolution and Strange Attractors}.
\newblock Lezioni Lincee. Accademia Nazionale dei Lincei (1989)

\bibitem{Rutgers1998}
Rutgers, M.A.: {Forced 2D turbulence: experimental evidence of simultaneous
  inverse energy and forward enstrophy cascades}.
\newblock Phys. Rev. Lett. \textbf{81}, 2244--2247 (1998)

\bibitem{SalmonBook}
Salmon, R.: {Lectures on Geophysical Fluid Dynamics}.
\newblock Oxford University Press (1998)

\bibitem{Serre1984}
Serre, D.: {Les invariants du premier ordre de l'{\'e}quation d'Euler en
  dimension trois}.
\newblock Physica D \textbf{13}, 105--136 (1984).
\newblock \doi{10.1016/0167-2789(84)90273-2}

\bibitem{LSmith1993}
Smith, L.M., Yakhot, V.: {Bose condensation and small-scale structure
  generation in a random force driven 2D turbulence}.
\newblock Phys. Rev. Lett. \textbf{71}, 352--355 (1993)

\bibitem{Sommeria2001b}
Sommeria, J.: {Course 8: Two-Dimensional Turbulence}.
\newblock In: M.~Lesieur, A.~Yaglom, F.~David (eds.) New Trends in Turbulence,
  Les Houches Theoretical Physics Summer School, chap.~8. Springer (2001)

\bibitem{Tabeling2002}
Tabeling, P.: Two-dimensional turbulence: a physicist approach.
\newblock Phys. Rep. \textbf{362}, 1--62 (2002)

\bibitem{Thalabard2014}
Thalabard, S., Dubrulle, B., Bouchet, F.: {Statistical mechanics of the 3D
  axisymmetric Euler equations in a Taylor-Couette geometry}.
\newblock J. Stat. Mech. p. P01005 (2014).
\newblock \urlprefix\url{http://arxiv.org/abs/1306.1081}

\bibitem{Touchette2009}
Touchette, H.: The large deviation approach to statistical mechanics.
\newblock Phys. Rep. \textbf{478}, 1--69 (2009).
\newblock \doi{10.1016/j.physrep.2009.05.002}

\bibitem{Touchette2004}
Touchette, H., Ellis, R.S., Turkington, B.: An introduction to the
  thermodynamic and macrostate levels of nonequivalent ensembles.
\newblock Physica A \textbf{340}, 138--146 (2004).
\newblock \doi{10.1016/j.physa.2004.03.088}

\bibitem{Turkington1999}
Turkington, B.: {Statistical Equilibrium Measures and Coherent States in
  Two-Dimensional Turbulence}.
\newblock Comm. Pure Appl. Math. \textbf{52}, 781--809 (1999).
\newblock \doi{10.1002/(SICI)1097-0312(199907)52:7<781::AID-CPA1>3.0.CO;2-C}

\bibitem{Turkington1996}
Turkington, B., Whitaker, N.: Statistical equilibrium computations of coherent
  structures in turbulent shear layers.
\newblock SIAM J. Sci. Comput. \textbf{17}, 1414 (1996).
\newblock \doi{10.1137/S1064827593251708}

\bibitem{VallisBook}
Vallis, G.K.: {Atmospheric and Oceanic Fluid Dynamics: Fundamentals and
  Large-scale Circulation}.
\newblock Cambridge University Press (2006)

\bibitem{Venaille2012b}
Venaille, A.: {Bottom-trapped currents as statistical equilibrium states above
  topographic anomalies}.
\newblock J. Fluid Mech. \textbf{699}, 500 (2012).
\newblock \doi{10.1017/jfm.2012.146}

\bibitem{Venaille2011b}
Venaille, A., Bouchet, F.: {Oceanic rings and jets as statistical equilibrium
  states}.
\newblock J. Phys. Oceanogr. \textbf{41}, 1860 (2011).
\newblock \doi{10.1175/2011JPO4583.1}

\bibitem{Venaille2012a}
Venaille, A., Vallis, G.K., Griffies, S.M.: {The catalytic role of beta effect
  in barotropization processes}.
\newblock J. Fluid Mech. \textbf{709}, 490--515 (2012).
\newblock \doi{10.1017/jfm.2012.344}

\end{thebibliography}

\end{document}